\documentclass[10pt,twocolumn,preprintnumbers,amsmath,amssymb,nofootinbib
,superscriptaddress]{revtex4-1}
\usepackage{graphicx,longtable,mathrsfs,color,array}
\usepackage[hidelinks]{hyperref}
\usepackage[usenames,dvipsnames]{xcolor} 
\usepackage{amssymb,amsmath,mathtools,mathrsfs,slashed} 
\usepackage{epsfig,subfigure,placeins,float} 
\usepackage{booktabs,longtable,ctable,multirow} 
\usepackage{exscale,relsize} 
\usepackage[normalem]{ulem} 
\usepackage{enumerate}
\usepackage{times, mathptmx} 
\usepackage[utf8]{inputenc}
\usepackage{color}
\usepackage{hyperref}
\usepackage{graphicx}
\usepackage{color}
\usepackage{graphicx,graphics}
\usepackage{bm}
\usepackage{tabularx}
\allowdisplaybreaks[1]

\begin{document}

\title{Quasi-normal modes of black holes in Horndeski gravity}
\author{Oliver J. Tattersall}
\email{oliver.tattersall@physics.ox.ac.uk}
\affiliation{Astrophysics, University of Oxford, DWB, Keble Road, Oxford OX1 3RH, UK}
\author{Pedro G. Ferreira}
\email{p.ferreira1@physics.ox.ac.uk}
\affiliation{Astrophysics, University of Oxford, DWB, Keble Road, Oxford OX1 3RH, UK}
\date{Received \today; published -- 00, 0000}

\begin{abstract}
We study the perturbations to General Relativistic black holes (i.e. those without scalar hair) in Horndeski scalar-tensor gravity. First, we derive the equations of odd and even parity perturbations of both the metric and scalar field in the case of a Schwarzschild black hole, and show that the gravitational waves emitted from such a system contain a mixture of quasi-normal mode frequencies from the usual General Relativistic spectrum and those from the new scalar field spectrum, with the new scalar spectrum characterised by just two free parameters. We then specialise to the sub-family of Horndeski theories in which gravitational waves propagate at the speed of light $c$ on cosmological backgrounds; the scalar quasi-normal mode spectrum of such theories is characterised by just a single parameter $\mu$ acting as an effective mass of the scalar field. Analytical expressions for the quasi-normal mode frequencies of the scalar spectrum in this sub-family of theories are provided for both static and slowly rotating black holes. In both regimes comparisons to quasi-normal modes calculated numerically show good agreement with those calculated analytically in this work.
\end{abstract}

\maketitle
\section{Introduction}

Einstein's theory of General Relativity (GR) is currently the best description of gravity available to us, having survived over 100 years of testing \cite{Will:2014kxa}. It is widely accepted to be the correct description of gravity at Solar System scales, where not only do its predictions show remarkable agreement with astrophysical data, but precise measurements of phenomena such as light deflection around the Sun and perihelion shift of Mercury (among others) rule out many modifications to GR. 

There are, however, compelling arguments that GR may require modification at both very high and very low energy scales. At high energies, unavoidable singularities arise during gravitational collapses and the so-called renormalization problem limits the analysis of quantum states; while on cosmological scales, GR relies on the as yet unexplained presence of `dark energy' in order to explain the observed accelerated expansion of the Universe \cite{Clifton20121}.

When considering modifications to GR, scalar-tensor theories are among the simplest extensions to GR available, and have been studied extensively in both the strong gravity and cosmological regimes. In addition, scalar-tensor theories appear to be ubiquitous in physics, as they appear as some limits of other theories of gravity, such as the decoupling limit of massive gravity \cite{deRham:2011by}. 

In this paper we will study the signatures that scalar-tensor modifications to GR could leave on the gravitational wave signal emitted from a perturbed black hole (for example, the ringdown signal of the remnant of a binary black hole merger \cite{Berti:2018vdi}). The frequencies of gravitational waves emitted from such a system are given by the quasi-normal modes (QNMs) of the remnant black hole, the eigenfrequencies of wave equations that arise when studying perturbations to the gravitational field equations on a black hole background \cite{Kokkotas:1999bd,0264-9381-16-12-201,Berti:2009kk}. Such frequencies depend on the theory of gravity considered, and so observations of the ringdown can provide a way to discriminate between GR and possible modified theories of gravity \cite{Berti:2005ys,Berti:2009kk,Dreyer:2003bv}. 

Given the recent advent of gravitational wave astronomy with numerous observations of mergers made by advanced LIGO and VIRGO \cite{2016PhRvL.116f1102A,2016PhRvL.116x1103A,2017PhRvL.118v1101A,Abbott:2017oio,PhysRevLett.119.161101}, in addition to the plans of future observatories such as eLISA, KAGRA, and the Einstein Telescope, there is no better time to study the physics of strong gravity systems in varying theories of gravity. Only with a full understanding of the possible signatures of modified theories of gravity that may arise can we fully take advantage of the remarkable observations currently being made (and those that have yet to be made). 

\textit{Summary}: In section \ref{secbackground} we will introduce the Horndeski family of scalar-tensor theories of gravity \cite{Horndeski:1974wa}, as well as the black hole backgrounds that are being considered in this paper. In sections \ref{secaction}-\ref{secEOM} we will calculate the quadratic action for perturbations to static black holes in Horndeski gravity and derive the equations of motion for both odd and even parity perturbations. We will show that the effect of Horndeski gravity modifications to the QNM frequencies of a perturbed black hole are characterised by just two free parameters. In section \ref{secQNM} we will derive analytical expressions for the QNMs of both static and slowly rotating black holes in a sub-family of Horndeski theories using the method of \cite{2009CQGra..26v5003D}, and provide comparisons to numerically calculated QNMs \cite{Konoplya:2004wg,2006PhRvD..73l4040K}. Finally, we will discuss the results of this paper and future work in section \ref{conclusion}. Throughout we will use natural units with $G=c=1$.

\section{Horndeski Gravity and background}\label{secbackground}

A general action for scalar-tensor gravity with 2$^{nd}$ order-derivative equations of motion is given by the Horndeski action \cite{Horndeski:1974wa}:
\begin{align}
S=\int d^4x\sqrt{-g}\sum_{n=2}^5L_n,\label{Shorndeski}
\end{align}
where the Horndeski Lagrangians are given by:
\begin{align}
L_2&=G_2(\phi,X)\nonumber\\
L_3&=-G_3(\phi,X)\Box \phi\nonumber\\
L_4&=G_4(\phi,X)R+G_{4X}(\phi,X)((\Box\phi)^2-\phi^{\mu\nu}\phi_{\mu\nu} )\nonumber\\
L_5&=G_5(\phi,X)G_{\mu\nu}\phi^{\mu\nu}-\frac{1}{6}G_{5X}(\phi,X)((\Box\phi)^3 \nonumber\\
& -3\phi^{\mu\nu}\phi_{\mu\nu}\Box\phi +2 \phi_{\mu\nu}\phi^{\mu\sigma}\phi^{\nu}_{\sigma}),
\end{align}
where $\phi$ is the scalar field with kinetic term $X=-\phi_\mu\phi^\mu/2$, $\phi_\mu=\nabla_\mu\phi$, $\phi_{\mu\nu}=\nabla_\nu\nabla_\mu\phi$, and $G_{\mu\nu}=R_{\mu\nu}-\frac{1}{2}R\,g_{\mu\nu}$ is the Einstein tensor. The $G_i$ denote arbitrary functions of $\phi$ and $X$, with derivatives $G_{iX}$ with respect to $X$. GR is given by the choice $G_4=M_{Pl}^2/2$ with all other $G_i$ vanishing and $M_{Pl}$ being the reduced Planck mass. Equation \ref{Shorndeski} is not the most general action and it has been shown that it can be extended to an arbitrary number of terms, exploiting degeneracies in the Hamiltonian \cite{Zumalacarregui:2013pma,Gleyzes:2014qga,Gleyzes:2014dya,Achour:2016rkg}.

For the background spacetime of the black hole, we assume a static, spherically symmetric ansatz for the line element and scalar field:
\begin{align}
ds^2=&-A(r)dt^2+\frac{1}{B(r)}dr^2+r^2\left(d\theta^2+\sin^2\theta d\phi^2\right)\label{metric}\\
\bar\phi=&\bar\phi(r).
\end{align}

No hair theorems exist for shift-symmetric theories (i.e. $G_{i\phi}=0$) \cite{Hui:2012qt,Sotiriou:2015pka,Maselli:2015yva} and generalised Brans-Dicke like theories ($G_3=G_{4X}=G_5=0$) \cite{PhysRev.124.925,Sotiriou:2015pka} such that $A=B=1-2M/r$ and $\bar\phi=\phi_0=\text{constant}$. For general $G_i$, various conditions can be found to ensure that the black hole solution is Schwarzschild with a constant scalar field profile \cite{2014PhRvD..89h4056G,PhysRevD.97.084005,Motohashi:2018wdq}. 

For the rest of this paper, we will assume that the black holes we are concerned with are indeed described by a Schwarzschild geometry with a constant background scalar field profile:
\begin{align}
ds^2=&-f(r)dt^2+\frac{1}{f(r)}dr^2+r^2\left(d\theta^2+\sin^2\theta d\phi^2\right)\label{schmetric}\\
\bar\phi=&\phi_0,\label{scalarbg}
\end{align}
with $f(r)=1-2M/r$. Whilst the background described by eq.~(\ref{schmetric}) is, in this case, identical to GR, in \cite{2018PhRvD..97d4021T} it was shown that the presence of additional degrees of freedom that can be excited (in this case a scalar field) can lead to non-GR signatures in the QNM spectrum of the black hole when it is perturbed. It is this QNM signature that we will proceed to investigate. 

\section{Quadratic action}\label{secaction}

To study the behaviour of perturbations about a background given by eq.~(\ref{schmetric})-(\ref{scalarbg}) in a theory described by the action given in eq.~(\ref{Shorndeski}), we introduce the perturbed fields $h_{\mu\nu}$ and $\delta\phi$, such that
\begin{align}
g_{\mu\nu}=&\;\bar{g}_{\mu\nu}+h_{\mu\nu}(x^\mu)\\
\phi=&\;\phi_0 + \delta\phi(x^\mu),
\end{align}
with $\bar{g}_{\mu\nu}$ being the Schwarzschild metric such that $\bar{g}_{\mu\nu} dx^\mu dx^\nu=ds^2$ as given in eq.~(\ref{schmetric}). Now, perturbing each term in eq.~(\ref{Shorndeski}) up to second order in the perturbed fields and collecting the terms quadratic in $h_{\mu\nu}$ and $\delta\phi$, we find the following quadratic action for perturbations:
\begin{widetext}
\begin{align}
S^{(2)}=\;\frac{M_{Pl}^2}{2}\int\,d^4x\,\sqrt{-g}&\left[\bar{G}_4\left(\frac{1}{4}\nabla_\mu h \nabla^\mu h +\frac{1}{2}\nabla_\alpha h_{\mu\nu}\nabla^\mu h^{\nu\alpha}-\frac{1}{4}\nabla_\alpha h_{\mu\nu} \nabla^\alpha h^{\mu\nu}-\frac{1}{2}\nabla_\mu h \nabla_\nu h^{\mu\nu}\right)+\left(\bar{G}_{3\phi}-\frac{1}{2}\bar{G}_{2X}\right)\nabla_\mu\delta\phi\nabla^\mu\delta\phi \right.\nonumber\\
&\left.+\bar{G}_{4\phi}\left(\nabla_\mu\delta\phi\nabla^\mu h -\nabla_\mu\delta\phi\nabla_\nu h^{\mu\nu}\right)+\bar{G}_{4X}\left(\left(\Box\delta\phi\right)^2-\nabla_\mu\nabla_\nu \delta\phi \nabla^\mu\nabla^\nu\delta\phi\right)+\frac{1}{2}\bar{G}_{2\phi\phi}\left(\delta\phi\right)^2\right],\label{Squad}
\end{align}
\end{widetext}
where an overbar indicates that the `barred' quantity should be evaluated at the background values of $g_{\mu\nu}$ and $\phi$. We have also factored out an overall factor of $M_{Pl}^2/2$, thus identifying $\bar{G}_4=1$ to obtain the correct GR limit.

From the background equations of motion \cite{Maselli:2015yva} we find that $\bar{G}_2=\bar{G}_{2\phi}=0$, whilst contributions from $G_5$ vanish due to the Bianchi identity. We further found that we had to impose $\bar{G}_3=0$ to ensure that the action is invariant under diffeomorphisms in this no-hair regime. Intuitively, these conditions make sense if we consider a simple model e.g. $G_2=X-V(\phi)$. If $\phi$ has a trivial background profile then $\bar{X}=\nabla_\mu \phi_0 \nabla^\mu \phi_0=0$, whilst $V(\phi_0)=0$ sets a `Cosmological Constant-like' term to zero. Furthermore, requiring that $G_{2\phi}=0$ is equivalent to requiring that $\phi$ sits at a minimum of the potential $V$. The additional constraint that $\bar{G}_3=0$ is unsurprising given that, to our knowledge, it has not been possible to formulate a no-hair theorem with the inclusion of a generic cubic term (with, of course, the exception of the shift-symmetric case \cite{Hui:2012qt,Sotiriou:2015pka,Maselli:2015yva}). An example of a non-zero $G_3$ that obeys this constraint is:
\begin{align}
G_3=\sum_{n=1}^{n=\infty}a_n \left(\phi-\phi_0\right)^n,
\end{align}
which satisfies $G_3(\phi_0)=0$, whilst giving $G_{3\phi}(\phi_0)=a_1$.

\section{Equations of motion}\label{secEOM}

We can take advantage of the spherical symmetry of the problem by separating the angular dependence of $h_{\mu\nu}$ and $\delta\phi$ from the time and radial dependence. The perturbation fields can be decomposed into tensor spherical harmonics, with the tensor perturbation $h_{\mu\nu}$ having both odd and even parity perturbations \cite{Regge:1957td,Rezzolla:2003ua}:
 \begin{align}
 h_{\mu\nu,\ell m}^{odd}=&
 \begin{pmatrix}
 0&0&h_0(r)B^{\ell m}_\theta&h_0(r)B^{\ell m}_\phi\\
 0&0&h_1(r)B^{\ell m}_\theta&h_1(r)B^{\ell m}_\phi\\
 sym&sym&0&0\\
 sym&sym&0&0
 \end{pmatrix}e^{-i\omega_{\ell m} t},
 \label{hodd}\\
 h_{\mu\nu,\ell m}^{even}=&
 \begin{pmatrix}
 H_0(r)f&H_1(r)&0&0\\
 sym&\frac{H_2(r)}{f}&0&0\\
 0&0&K(r)r^2&0\\
 0&0&0&K(r)r^2\sin\theta
 \end{pmatrix}Y^{\ell m}e^{-i\omega_{\ell m} t},
 \label{heven}
\end{align}
whilst the scalar perturbation $\delta\phi$ is purely of even parity:
\begin{align}
\delta\phi_{\ell m}=\frac{\varphi(r)}{r}e^{-i\omega_{\ell m} t}Y^{\ell m},
\end{align}
where $sym$ indicates a symmetric entry, $B^{\ell m}_\mu$ is the odd parity vector spherical harmonic and $Y^{\ell m}$ is the standard scalar spherical harmonic, as described in \cite{Martel:2005ir,Ripley:2017kqg} (note there are slight differences in convention between the definitions of tensorial spherical harmonics used in \cite{Martel:2005ir} and \cite{Ripley:2017kqg}). Furthermore, we are working in the Regge-Wheeler gauge for simplicity \cite{Regge:1957td}. Thus, the amplitude of linear perturbations is described by the functions $h_i$, $H_i$, $K$, and $\varphi$. We have further assumed a time dependence of $e^{-i\omega_{\ell m} t}$ for our perturbations, due to the static nature of the background spacetime. It is these frequencies $\omega_{\ell m}$ that will prove to be the QNMs of the perturbed black hole. Spherical harmonic indices will be omitted from now on, with each equation assumed to hold for a given $\ell$ (we will find that the equations of motion are independent of $m$, which is unsurprising due to the spherical symmetry of the background). In general, the perturbed fields will be represented by a sum over $\ell$, $m$, and $\omega$ of the modes. The quadratic actions of both odd and even parity perturbations in Horndeski gravity on generic spherically symmetric backgrounds were studied in \cite{Kobayashi:2012kh,Kobayashi:2014wsa}.

\subsection{Odd Parity}

Upon varying the action given by eq.~(\ref{Squad}), we find that the odd parity metric perturbations are governed by the standard Regge-Wheeler equation exactly as in GR \cite{Regge:1957td,2018PhRvD..97d4021T}. This is, of course, unsurprising given that the background spacetime is given by the GR solution, and that the even parity scalar field perturbations do not couple to the odd parity metric perturbations. The spectrum of QNMs arising from solving the Regge-Wheeler equation is well known \cite{1975RSPSA.344..441C,Kokkotas:1999bd,0264-9381-16-12-201,Berti:2009kk}.

\subsection{Even Parity}

Upon varying the action given by eq.~(\ref{Squad}), we find that the even parity metric and scalar perturbations are governed by two homogeneous second order wave equations. We find that a `Zerilli-like' function $\Psi$ satisfies the standard GR Zerilli equation \cite{Zerilli:1970se,PhysRevD.34.333,2018PhRvD..97d4021T}:
\begin{align}
\frac{d^2\Psi}{dr_\ast^2}+ \left[\omega^2-V_Z(r)\right]\Psi=0,
\end{align}
with
\begin{align}
V_Z(r)=&\;2\left(1-\frac{2M}{r}\right)\frac{\Lambda^2r^2\left[(\Lambda+1)r+3M\right]+9M^2(\Lambda r+M)}{r^3(\Lambda r+3M)^2}\nonumber\\
2\Lambda=&\;(\ell+2)(\ell-1),
\end{align}
and the `tortoise coordinate' $r_\ast$ satisfying $dr_\ast = f(r)^{-1}dr$. The scalar perturbation $\varphi$, meanwhile, satisfies:
\begin{align}
\frac{d^2\varphi}{dr_\ast^2}+\left[\omega^2-V_S(r)\right]\varphi=0,\label{scalarEOM}
\end{align}
with
\begin{align}
V_S(r)=&\left(1-\frac{2M}{r}\right)\left(\frac{2M}{r^3}-\frac{\bar{G}_{2\phi\phi}}{3\bar{G}_{4\phi}^2+\bar{G}_{2X}-2\bar{G}_{3\phi}}\right.\nonumber\\
&\left.+\frac{\ell(\ell+1)}{r^2}\left[1+\frac{8\bar{G}_{4X}(1-2M/r)}{3\bar{G}_{4\phi}^2+\bar{G}_{2X}-2\bar{G}_{3\phi}}\right]\right).\label{vscalar}
\end{align}
The even parity metric functions $H_0, H_1, H_2,$ and $K$ are given found in terms of $\Psi$ and $\varphi$ to be:
\begin{align}
H_0=&\frac{\partial}{\partial r}\left[\left(1-\frac{2M}{r}\right)\left(g_2(r)\Psi+r\frac{\partial \Psi}{\partial r}\right)\right]-K\\
H_1=&-i\omega \left(g_2(r)\Psi+r\frac{\partial \Psi}{\partial r}\right)\\
H_2=&H_0-2\bar{G}_{4\phi}\frac{\varphi}{r}\label{H2def}\\
K=&g_1(r)\Psi+\left(1-\frac{2M}{r}\right)\frac{\partial \Psi}{\partial r}-\bar{G}_{4\phi}\frac{\varphi}{r},\label{Kdef}
\end{align}
with
\begin{align}
g_1(r)=&\frac{\Lambda(\Lambda+1)r^2+2\Lambda Mr+6M^2}{r^2(\Lambda r+3M)},\nonumber\\
g_2(r)=&\frac{\Lambda r^2-3\Lambda Mr-3M^2}{(r-2M)(\Lambda r+3M)}.
\end{align}

Note that the above set of equations show a mixing between the perturbed metric components and the perturbed scalar field, e.g. in eq.~(\ref{H2def})-(\ref{Kdef}). Thus, whilst $\Psi$ satisfies the GR Zerilli equation (with accompanying GR spectrum of QNMs), the metric perturbation $h_{\mu\nu}$ will contain a mix of GR \textit{and} scalar modes. We can make this clear by writing the even parity metric perturbation for Horndeski gravity as:
\begin{align}
h_{\mu\nu,\ell m}=h_{\mu\nu,\ell m}^{GR}-\bar{G}_{4\phi}\bar{g}_{\mu\nu}\delta\phi_{\ell m}Y^{\ell m}e^{-i\omega_{\ell m} t},\label{eqmixing}
\end{align}
where $h_{\mu\nu,\ell m}^{GR}$ is the standard even parity metric perturbation one would calculate for a Schwarzschild black hole in GR. Clearly, the QNMs of gravitational waves emitted from such a perturbed black hole will therefore exhibit a mixture of those frequencies arising from the standard GR spectrum \textit{and the modified scalar spectrum}. This is one of the main results of this section (and was first described in \cite{2018PhRvD..97d4021T}) . 

Eq.~(\ref{eqmixing}) clearly shows that the mixing of metric and scalar perturbations is due to the conformal coupling between $\phi$ and curvature. Thus in theories with $\bar{G}_{4\phi}=0$ the metric and scalar perturbations will decouple (see eq.~(\ref{H2def})-(\ref{Kdef})), leaving no modified gravity signature in the metric perturbation. For example, in Einstein-Scalar-Gauss-Bonnet (ESGB) gravity, the scalar field perturbations obey a massive Klein-Gordon equation when the background spacetime is given by a GR black hole \cite{Silva:2017uqg}. When recast into the language of Horndeski theories, however, $\bar{G}_{4\phi}=0$ for ESGB gravity (see e.g. \cite{Gong:2017kim}). Perturbed GR black holes in ESGB gravity will, therefore, not exhibit a mixing of GR and scalar QNMs in the emission of gravitational waves (assuming a Schwarzschild geometry with constant scalar field profile for the background). 

Eq.~(\ref{vscalar}) shows that the scalar spectrum is characterised by two parameters dependent on the free functions present in the Horndeski Lagrangian:
\begin{align}
\mu^2=\frac{-\bar{G}_{2\phi\phi}}{3\bar{G}_{4\phi}^2+\bar{G}_{2X}-2\bar{G}_{3\phi}},\;\Gamma=\frac{8\bar{G}_{4X}}{3\bar{G}_{4\phi}^2+\bar{G}_{2X}-2\bar{G}_{3\phi}}.\label{params}
\end{align}
In terms of these new parameters, the scalar potential $V_S(r)$ takes the form:
\begin{align}
V_S(r)=&\left(1-\frac{2M}{r}\right)\left(\frac{\ell(\ell+1)}{r^2}\left[1+\Gamma\left(1-\frac{2M}{r}\right)\right]\right.\nonumber\\
&\left.+\frac{2M}{r^3}+\mu^2\right).\label{vscalar2}
\end{align} 
The identification of these two combinations of the $G_i$ and their derivatives as the only parameters characterising the scalar QNM spectrum is the second main result of this section. If $\mu=\Gamma=0$ then eq.~(\ref{scalarEOM}) takes the form of a massless Klein-Gordon equation on the Schwarzschild background. 

\section{Quasinormal modes}\label{secQNM}

We will now focus on the sub-family of Horndeski theories in which gravitational waves propagate at the speed of light on cosmological backgrounds (and where the scalar field has a non-negligible impact on the cosmology), as is indicated by the detection of GW170817  and its optical counterpart GRB170817 \cite{PhysRevLett.119.161101,2041-8205-848-2-L12,2041-8205-848-2-L13,2041-8205-848-2-L14,2041-8205-848-2-L15,2017Sci...358.1556C}. It was shown in \cite{Lombriser:2015sxa,Lombriser:2016yzn,2017PhRvL.119y1301B,Creminelli:2017sry,Sakstein:2017xjx,Ezquiaga:2017ekz} that this restriction on the speed of gravitational waves $c_T$ gives the following constraints on the Horndeski parameters:
\begin{align}
G_{4X}=G_{5X}=G_{5\phi}=0.\label{constraints}
\end{align}
The resulting constrained Horndeski action is thus given by:
\begin{align}
S=\frac{M_{Pl}^2}{2}\int d^4x \,\sqrt{-g}\,\left[\phi R+G_2(\phi,X)-G_3(\phi,X)\Box\phi\right],\label{Sgen}
\end{align}
where we have made field redefinitions to set $G_4=\phi$ and taken out an overall pre-factor of one half the reduced Planck mass, $M_{Pl}^2/2$, thus setting $G_{4\phi}=\phi_0=1$; the quintic Horndeski term with constant $G_5$ vanishes due to the Bianchi identity. In \cite{PhysRevD.97.084005} it was shown that in such theories where the non-GR degrees of freedom have cosmological relevance, the constraint that $c_T=1$ leads in many cases to black holes without hair (as we have considered so far in this paper). 

The action given by eq.~(\ref{Sgen}) is in the form of a conformally coupled scalar-tensor theory, with scalar potential and kinetic terms given by $G_2$ and an additional scalar cubic term given by $G_3$. This action can be transformed from the `Jordan' frame (in its current state) into the `Einstein' frame by making a conformal transformation \cite{Bettoni:2013diz}. In the Einstein frame the theory will take the form of GR with a minimally coupled scalar field, though any matter fields would now couple to a different metric than that which contributes the Einstein-Hilbert term in the action. For convenience we will continue to work in the Jordan frame as it is the Jordan frame metric that gravitational wave detectors (made, of course, of matter) will couple to \cite{Yunes:2013dva}.

Thus from now on we will set $\Gamma=0$ when using the results of section \ref{secEOM} to reflect the constraints of eq.~(\ref{constraints}), leaving the scalar effective mass $\mu$ as the only free parameter left characterising the scalar spectrum. 

\subsection{Static black holes}

For Schwarzschild black holes, the spectrum of QNM frequencies $\omega$ associated with the Regge-Wheeler and Zerilli equations (i.e. the GR spectrum) are well known and have been calculated with a variety of different methods \cite{Kokkotas:1999bd,0264-9381-16-12-201,Berti:2009kk}. Thus we will focus on the spectrum arising from the scalar equation of motion (remembering that, as explained in the previous section, gravitational waves will contain a mixture of modes from both spectra). With $\Gamma=0$, eq.~(\ref{scalarEOM}) is in the form of a massive Klein-Gordon equation:
\begin{align}
(\Box-\mu^2)\delta\phi=0,
\end{align}
leading to the interpretation of $\mu^2$  as an effective mass given by eq.~(\ref{params}). The same effective mass for the scalar field is found when linearising Horndeski gravity about a flat background \cite{Gong:2017bru}. The QNMs of a massive scalar field on a black hole background have been calculated numerically in \cite{Konoplya:2004wg,2006PhRvD..73l4040K}. To show the explicit dependence of the frequencies on $\mu$, however, we present here an analytical expansion in $L=\ell+1/2$ of the scalar QNMs using the technique developed in \cite{2009CQGra..26v5003D}, with $\ell$ being the multi-polar spherical harmonic index:
\begin{align}
\omega=\sum_{k=-1}^{k=\infty}\omega_k \;L^{-k}.\label{expansion}
\end{align}
The details of the expansion, and its limitations, are discussed in detail in \cite{2009CQGra..26v5003D}. For the scalar perturbation equation of motion given by eq.~(\ref{scalarEOM}), the expansion coefficients $\omega_k$ are given by, up to $O(L^{-6})$:
\begin{widetext}
\begin{align}
\sqrt{27}M\omega_{-1}=&1\nonumber\\
\sqrt{27}M\omega_0=&-iN\nonumber\\
\sqrt{27}M\omega_1=&\frac{29}{432}-\frac{5 N^2}{36}+\frac{9}{2}\mu ^2 M^2\nonumber\\
\sqrt{27}M\omega_2=&iN\left(-\frac{313}{15552}-\frac{235 N^2}{3888}+\frac{15}{2} \mu ^2 M^2\right)\nonumber\\
\sqrt{27}M\omega_3=&\frac{854160
   N^4+450312 N^2-82283}{40310784}-\frac{1}{288} \mu ^2 M^2 \left(2460 N^2+143\right)+\frac{27}{8}\mu^4 M^4\nonumber\\
\sqrt{27}M\omega_4=&\frac{i N}{2902376448} \left(11273136
   N^4+15675000 N^2+4832407+29386561536 \mu ^4 M^4-839808 \mu ^2 M^2 \left(27260 N^2+3893\right)\right)\nonumber\\
\sqrt{27}M\omega_5=&\frac{6 N^2 \left(99340528
   N^4-70621200 N^2-49716689\right)-248844479}{313456656384}+\frac{
  167490960 N^4+41998920 N^2+2237653}{26873856}\mu ^2 M^2\nonumber\\
   &-\frac{1}{384} \mu ^4 M^4 \left(1956 N^2+365\right)+\frac{189}{16}\mu ^6 M^6\nonumber\\
\sqrt{27}M\omega_6=&\frac{i N}{135413275557888} \left(347667122880 N^6+90232249296 N^4-50499755276
   N^2+356260748667\right.\nonumber\\
   &\left.+69984 \mu ^2 M^2 \left(8311972368 N^4+3193691880
   N^2+288558197\right)+29386561536 \mu ^4 M^4 \left(128740
   N^2-5933\right)\right.\nonumber\\
   &\left.+8454866392645632 \mu ^6 M^6\right),\label{omegasch}
\end{align}
\end{widetext}
where $N=n+1/2$ with $n$ being the overtone number. Whilst only the first eight terms of the expansion are given here, as discussed in \cite{2009CQGra..26v5003D}, one can readily extend calculate subsequent terms to an arbitrary order in $L$ with the use of a computer algebra package. The expansion coefficients provided in eq.~(\ref{omegasch}) are the main result of this section, providing an easy to use analytical expression for calculating QNMs for arbitrary effective scalar mass $\mu$, multi-polar index $\ell$, and overtone $n$ (with some limitations as discussed below and in  \cite{2009CQGra..26v5003D}). 

The expansion given by eq.~(\ref{expansion})-(\ref{omegasch}) is clearly more accurate for large values of $L$ and as such should not be used for $L<1$ i.e. $\ell=0$. Furthermore, as noted in \cite{2009CQGra..26v5003D}, the expansion is only accurate for $\ell > n$. Table \ref{table1} shows a comparison between the frequencies calculated in \cite{2006PhRvD..73l4040K} and those calculated using the above expansion for $\ell=1, n=0$. For the values of $\mu$ considered, the errors between the frequencies calculated using the expansion in $L$ and those calculated numerically never exceeds 1\%. This is an impressive level of agreement given the relatively small number of terms in the analytical expansion calculated here. Greater accuracy can be expected by calculated term to higher (inverse) order in $L$. 
\begin{figure}
\caption{Real and imaginary frequency components for the fundamental ($n=0$) mode as a function of the scalar effective mass $\mu M$ for different values of $\ell$.}
\label{fig1}
\includegraphics[width=0.5\textwidth]{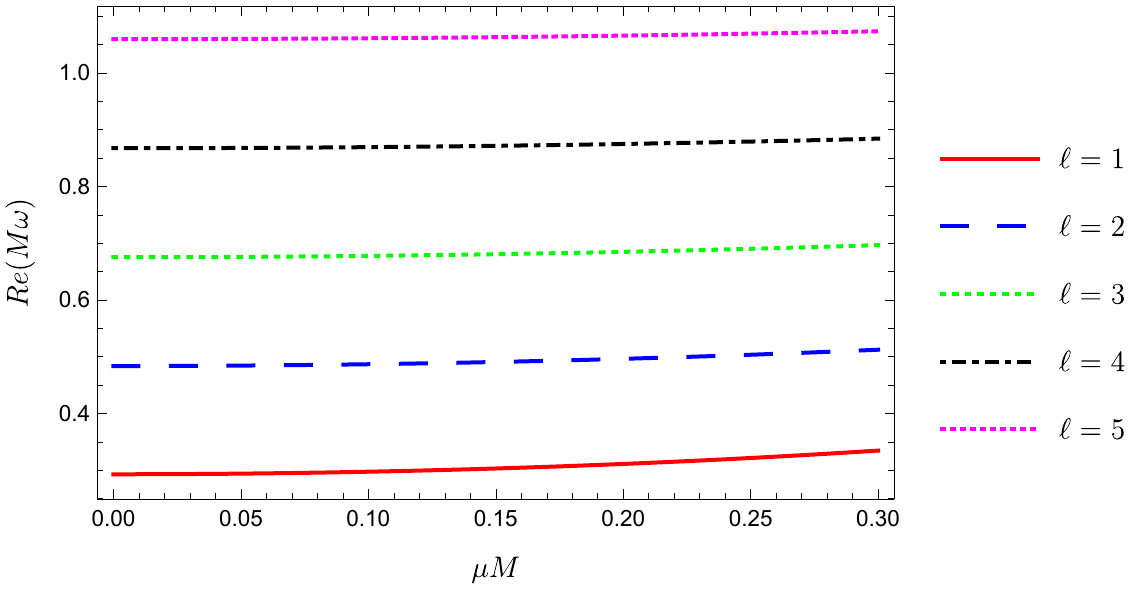}\\
\includegraphics[width=0.5\textwidth]{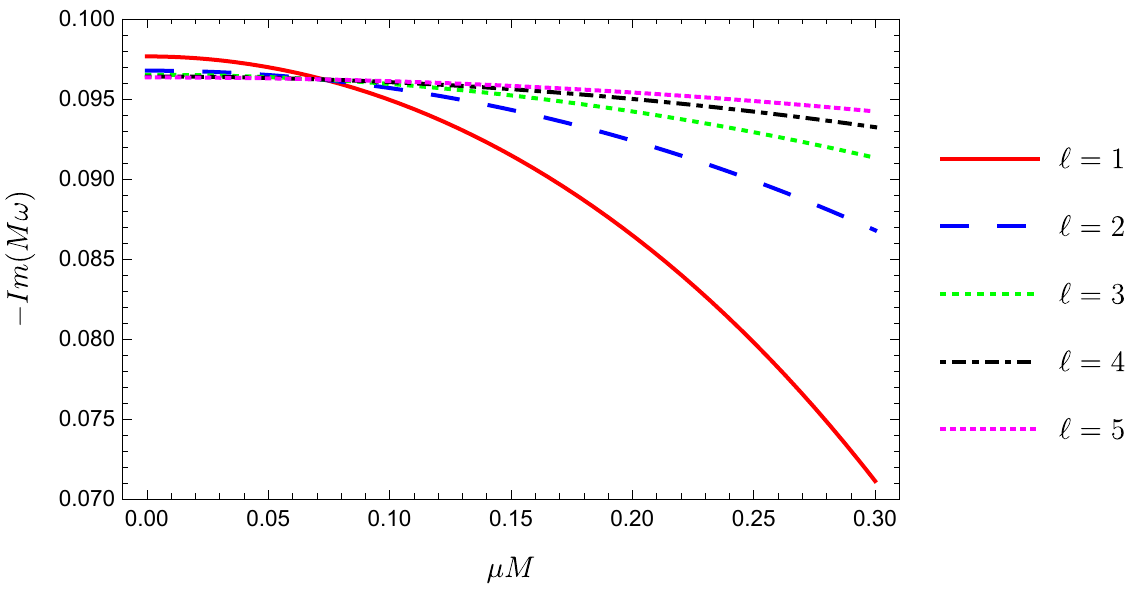}
\end{figure}

\begin{figure}
\caption{Complex QNMs for the fundamental ($n=0$) mode as a function of the scalar effective mass $\mu$ for different values of $\ell$ (the mass labels are suppressed for $\ell\neq2$).}
\label{fig5}
\includegraphics[width=0.5\textwidth]{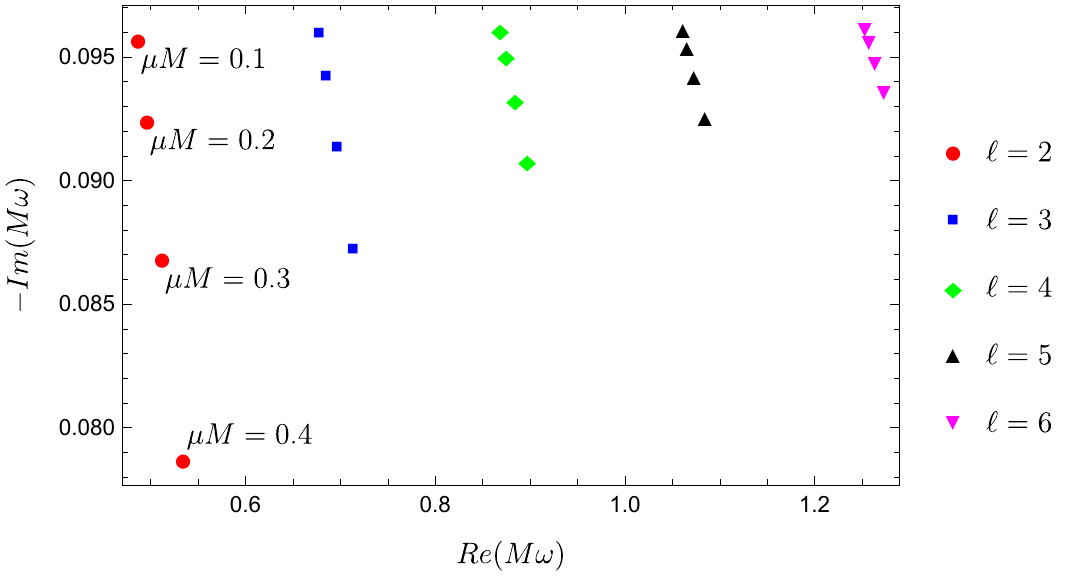}
\end{figure}

Figure \ref{fig1} shows the real and imaginary frequency components for the $n=0$ mode for different values of $\ell$ and $\mu$. All of the frequencies found have negative imaginary components, representing stable damped modes. The frequencies are progressively less damped for larger values of the scalar effective mass $\mu$. Figure \ref{fig5} shows the migration of the QNMs through the complex plane for different values of $\mu$, with the magnitude of the damping frequency decreasing with increasing effective mass $\mu$. 

The expression for QNMs given by eq.~(\ref{omegasch}) should be used with caution for values of $\mu$ larger than those used in this paper (around $\mu M=0.3$). For example, the $\ell=1,n=0$ mode calculated using eq.~(\ref{omegasch}) with $\mu M=0.53$ is $\omega_{1,0}=0.44+0.0086i$, i.e. an unstable mode, whilst \cite{Konoplya:2004wg} shows the same mode having a negative (and therefore stable) imaginary component. We suspect that this inaccuracy is due to our relatively early truncation of the series in $L$ (only the first 8 terms of the series have been provided here). As can be seen in eq.~(\ref{omegasch}), the expansion coefficients $\omega_k$ contain higher powers of $\mu M$ as $k$ increases, thus higher order $\omega_k$ become more relevant as the value of $\mu M$ increases. If one wishes to to calculate QNMs accurately for larger values of the scalar effective mass, a greater number of terms will need to be calculated in the expansion. This kind of limitation is unsurprising given that we have only calculated terms to $O(L^{-6})$; in \cite{2009CQGra..26v5003D} expressions up to $O(L^{-12})$ and beyond are calculated with relative ease.

\begin{table*}
\caption{Comparison of the quasinormal frequencies calculated by numerical and analytic expansion techniques for $\ell=1,n=0$ for different values of effective mass $\mu$.}
\label{table1}
\begin{ruledtabular}
\begin{tabular}{ccccccc}
\multicolumn{1}{l}{} &
\multicolumn{2}{c}{Numerical}&
\multicolumn{2}{c}{L-expansion}&
\multicolumn{2}{c}{\% error}\\
$\mu M$ & Re($M\omega$) & -Im($M\omega$)  & Re($M\omega$) & -Im($M\omega$) & Re($M\omega$) & -Im($M\omega$)
\\
\hline
\\
0 &   0.292936 & 0.097660   & 0.292924 & 0.097649  & -0.004096   & -0.012390  \\
0.1 & 0.297416  &  0.094957  & 0.297429 &  0.094935 &  0.004371  & -0.023168  \\
0.2 & 0.310957 & 0.086593   & 0.311127 &  0.086474 &  0.054670  & 0.137613 \\
0.3 & 0.333777 & 0.071658  & 0.334668   & 0.071116  & 0.266945   &  -0.756371   \\
\end{tabular}
\end{ruledtabular}
\end{table*}

\subsection{Slowly rotating black holes}\label{secSR}

In general, astrophysical black holes are not expected to be static and spherically symmetric, but instead axisymmetric due to their possessing angular momentum \cite{1972CMaPh..25..152H}. As such, axisymmetric black holes are the objects expected to be involved in the events detected at gravitational wave observatories (and indeed, already have been at advanced LIGO and VIRGO \cite{2016PhRvL.116f1102A,2016PhRvL.116x1103A,2017PhRvL.118v1101A,Abbott:2017oio}). The calculation of QNMs that might arise from rotating black holes is therefore of great importance if theoretical predictions are to be tested with observations. In this section we will address slowly rotating black holes, i.e. those black holes for which their dimensionless angular momentum $a$ satisfies $|a| \ll 1$. 

In many cases, it has been shown that no-hair theorems that are valid in the static regime can be extended to apply also to slowly rotating black holes \cite{Hui:2012qt,Sotiriou:2015pka,Maselli:2015yva}. We will again assume that, in the slowly rotating regime, the background black hole in the theory given by eq.~(\ref{Shorndeski}) with the constraints given by eq.~(\ref{constraints}) is described by the corresponding GR solution (in this case the Kerr metric up to linear order in $a$) with a constant scalar field profile:
\begin{align}
ds^2=&-f(r)dt^2+\frac{1}{f(r)}dr^2+r^2\left(d\theta^2+\sin^2\theta d\phi^2\right)\nonumber\\
&-\frac{4M^2a}{r}\sin^2\theta dtd\phi\\
\phi(r,\theta)=&\;\phi_0.
\end{align}

We further expect that the gravitational waves emitted from such a black hole when perturbed are going to be a mixture of frequencies from the usual GR spectrum and those from the scalar field spectrum, as in the spherically symmetric case, due to the conformal coupling between $\phi$ and the metric. As the GR spectrum is, once again, well studied \cite{1985RSPSA.402..285L,Kokkotas:1999bd,0264-9381-16-12-201,Berti:2009kk}, we will focus on the scalar spectrum. 

In general, in axisymmetric (but non-spherically symmetric) spacetimes perturbations of different parity and $\ell$ mix, complicating the ability to separate out the angular and radial dependencies of the perturbed fields \cite{1973ApJ...185..635T}. It was, however, shown in \cite{2012PhRvD..86j4017P} that to calculate the QNM spectrum to first order in $a$, it is sufficient to consider that perturbations of different parity and $\ell$ \textit{do not} mix. In \cite{2012PhRvD..86j4017P} the equation of motion for massive scalar field perturbations about a Kerr background to \textit{first order in rotation} was shown to be:
\begin{align}
\frac{d^2\varphi}{dr_\ast^2}+\left[\omega^2-V_S(r)-\frac{4M^2\omega}{r^3}am\right]\varphi=0,\label{scalarEOMSR}
\end{align}
where $m$ is the azimuthal spherical harmonic index and $V_S(r)$ is given by eq.~(\ref{vscalar2}) with $\Gamma=0$. We see that the effect of a slowly rotating background is to introduce an additional spin-dependent term to the potential.

The QNMs of a massive scalar field in a Kerr background were calculated numerically in \cite{2006PhRvD..73l4040K}. In parallel to our analysis of the static perturbations, we will present an (in this case somewhat more crude) analytical expansion for the scalar frequencies using the technique of \cite{2009CQGra..26v5003D}. The technique developed in \cite{2009CQGra..26v5003D} has a geometric grounding, with the position of the light sphere at $r=3M$ playing a key role in calculating the expansion coefficients. With a slowly rotating black hole, the radii of circular photon orbits are instead given by $r_{\pm}=3M(1\pm2\sqrt{3}a/9$) \cite{2003GReGr..35.1909T}, thus it is reasonable to expect that an expansion based around $r=3M$ being a `privileged' position will not give especially accurate results. Nonetheless, in an effort to calculate an analytical expression to study the effect of the effective mass $\mu$ on the QNMs, the following expansion in $L$ is provided:
\begin{align}
\omega=\sum_{k=-1}^{k=\infty}\omega_k \;L^{-k},\label{expansion2}
\end{align}
with, again up to $O(L^{-6})$ and up to linear order in spin $a$:
\begin{widetext}
\begin{align}
\sqrt{27}M\omega_{-1}=&1\nonumber\\
\sqrt{27}M\omega_0=&-iN+\frac{2}{3\sqrt{3}}am\nonumber\\
\sqrt{27}M\omega_1=&\frac{29}{432}-\frac{5 N^2}{36}+\frac{9}{2}\mu ^2 M^2\nonumber\\
\sqrt{27}M\omega_2=&iN\left(-\frac{313}{15552}-\frac{235 N^2}{3888}+\frac{15}{2} \mu ^2 M^2\right)+\frac{am}{\sqrt{3}}\left(\frac{7}{324}+\frac{5}{27}N^2-6\mu^2M^2\right)\nonumber\\
\sqrt{27}M\omega_3=&\frac{854160
   N^4+450312 N^2-82283}{40310784}-\frac{1}{288} \mu ^2 M^2 \left(2460 N^2+143\right)+\frac{27}{8}\mu^4 M^4+\frac{iamN}{\sqrt{3}}\left(-\frac{29}{5832}+\frac{145}{1458}N^2-14\mu^2M^2\right)\nonumber\\
\sqrt{27}M\omega_4=&\frac{i N}{2902376448} \left(11273136
   N^4+15675000 N^2+4832407+29386561536 \mu ^4 M^4-839808 \mu ^2 M^2 \left(27260 N^2+3893\right)\right)\nonumber\\
   &-\frac{am}{\sqrt{3}}\left(\frac{44203}{10077696}+\frac{11329}{419904}N^2+\frac{6005}{209952}N^4-\left[\frac{5}{6}+20N^2\right]\mu^2M^2\right)\nonumber\\
\sqrt{27}M\omega_5=&\frac{6 N^2 \left(99340528
   N^4-70621200 N^2-49716689\right)-248844479}{313456656384}+\frac{
  167490960 N^4+41998920 N^2+2237653}{26873856}\mu ^2 M^2\nonumber\\
   &-\frac{1}{384} \mu ^4 M^4 \left(1956 N^2+365\right)+\frac{189}{16}\mu ^6 M^6+\frac{iamN}{\sqrt{3}}\left(\frac{1739663}{544195584}-\frac{221683}{22674816}N^2+\frac{86065}{11337408}\right.\nonumber\\
   &\left.+\left[\frac{2239}{864}+\frac{4645}{216}N^2\right]\mu^2 M^2+18\mu^4 M^4\right)\nonumber\\
\sqrt{27}M\omega_6=&\frac{i N}{135413275557888} \left(347667122880 N^6+90232249296 N^4-50499755276
   N^2+356260748667\right.\nonumber\\
   &\left.+69984 \mu ^2 M^2 \left(8311972368 N^4+3193691880
   N^2+288558197\right)+29386561536 \mu ^4 M^4 \left(128740
   N^2-5933\right)\right.\nonumber\\
   &\left.+8454866392645632 \mu ^6 M^6\right)+\frac{am}{\sqrt{3}}\left(\frac{66762331}{235092492288
  }+\frac{111853717
  N^2}{39182082048}-\frac{445955
   N^4}{272097792}-\frac{42152075 N^6}{2448880128}\right.\nonumber\\
   &\left.-\left[\frac{892021}{10077696}+\frac{1785679}{419904}N^2+\frac{394255}{209952}N^4\right]\mu^2M^2-\left[\frac{29}{18}+\frac{350}{3}N^2\right]\mu^4 M^4-18\mu^6M^6\right).\label{omegaSR}
\end{align}
\end{widetext}

\begin{table*}
\caption{Comparison of the slow-rotation quasi-normal frequencies calculated by numerical and analytic expansion techniques for $\ell=m=1,n=0$ for different values of effective mass $\mu$ and dimensionless angular momentum $a$.}
\label{table2}
\begin{ruledtabular}
\begin{tabular}{ccccccc}
\multicolumn{1}{l}{$a=0.1$} &
\multicolumn{2}{c}{Numerical}&
\multicolumn{2}{c}{L-expansion}&
\multicolumn{2}{c}{\% error}\\
$\mu M$ & Re($M\omega$) & -Im($M\omega$)  & Re($M\omega$) & -Im($M\omega$) & Re($M\omega$) & -Im($M\omega$)
\\
\hline
\\
0 & 0.301045   & 0.097547   & 0.300639 & 0.097614  & -0.134864   & 0.068685  \\
0.1 & 0.305329  &  0.095029  & 0.304950 &  0.095072 & -0.124128   & 0.045249  \\
0.2 &   0.318274  &  0.087228 & 0.318029  &  0.087108  & -0.076978 & -0.137571\\
\hline
 $a=0.2$ &    &    &  &   &    &   \\
$\mu M$ & Re($M\omega$) & -Im($M\omega$)  & Re($M\omega$) & -Im($M\omega$) & Re($M\omega$) & -Im($M\omega$)
\\
\hline
\\
0 &  0.310043   &  0.097245   &  0.308354& 0.097581  & -0.544763   &  0.345519 \\
0.1 &  0.314119  &   0.094920  & 0.312471 & 0.095209  &  -0.524642  & 0.304467  \\
0.2 &0.326433  & 0.087709   &0.324931  & 0.087742  & -0.460125   &  0.037624\\
\hline
 $a=0.3$ &    &    &  &   &    &   \\
$\mu M$ & Re($M\omega$) & -Im($M\omega$)  & Re($M\omega$) & -Im($M\omega$) & Re($M\omega$) & -Im($M\omega$)
\\
\hline
\\
0 &  0.320126  &  0.096691  & 0.316069 &  0.097547 &  -1.26731  & 0.885294  \\
0.1 & 0.323981  & 0.094569   & 0.319992 & 0.095346  & -1.23125   & 0.821622  \\
0.2 & 0.335621 &  0.087979  &0.331833  & 0.088376  & -1.12865   & 0.451244  \\
\end{tabular}
\end{ruledtabular}
\end{table*}

\begin{figure}
\caption{Percentage error of real and imaginary frequency components for the $\ell=m=1,n=0$ mode as a function of dimensionless angular momentum $a$ for different values of $\mu M$.}
\label{fig3}
\includegraphics[width=0.5\textwidth]{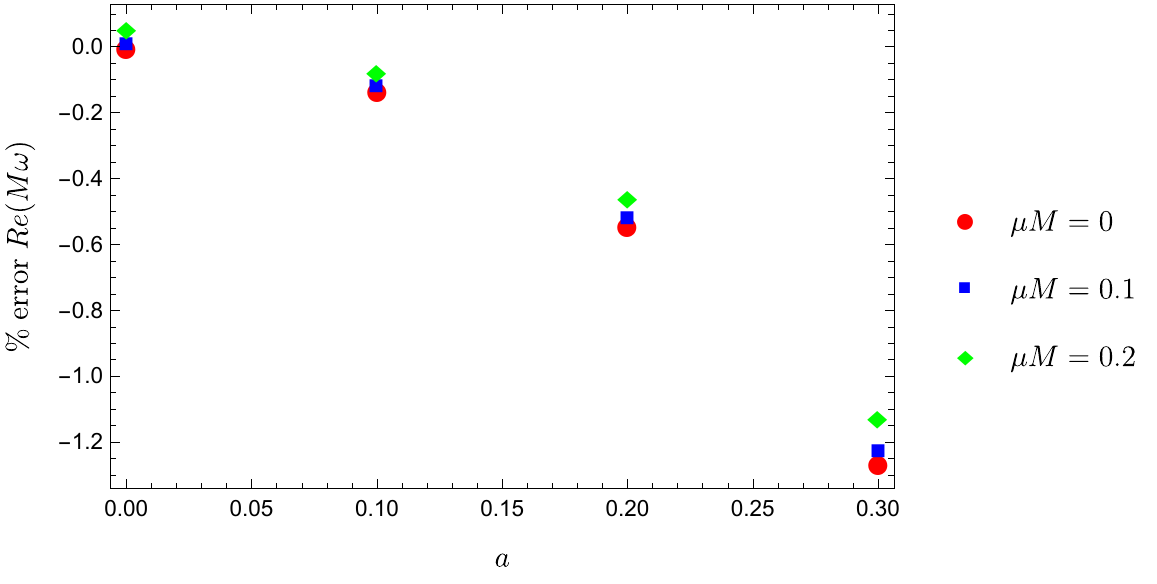}\\
\includegraphics[width=0.5\textwidth]{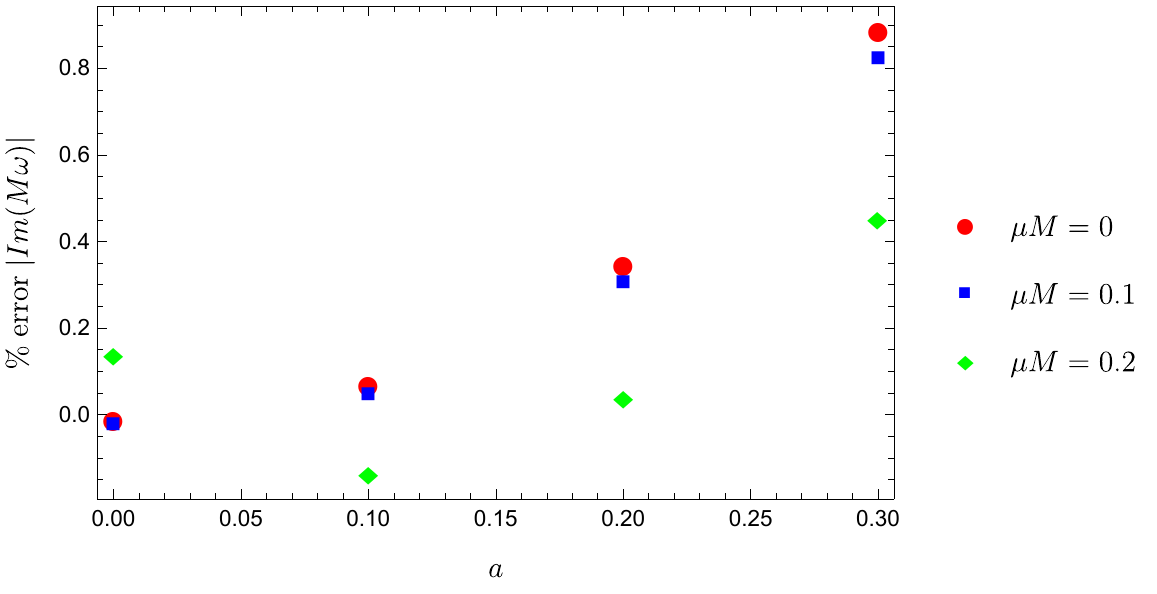}
\end{figure}

Table \ref{table2} gives a comparison of the frequencies calculated from the above expansion to those calculated numerically in \cite{2006PhRvD..73l4040K} for the fundamental $\ell=m=1$ mode. As expected, the errors between the two methods increase with increasing $a$, however up to $a=0.2$ the errors stay below 1\%. The error between the frequencies calculated numerically and those calculated with the analytical expansion provided in this paper decrease with increasing effective scalar mass $\mu$ for the imaginary frequency components, whilst a mild increase in error with increasing mass is seen for the real frequency components. Figure \ref{fig3} shows the percentage error between the two methods plotted as a function of dimensionless angular momentum $a$. Figure $\ref{fig3}$ shows a non-linear dependence on $a$ for the magnitude of the errors between the numerical and analytical methods, as expected given that the analytical method used here neglects terms $O(a^2)$ and above due to treating $a$ as a small parameter. We can attempt to fit a power law dependence of the magnitude of the errors on $a$, finding the following best fits (ignoring any constant offsets):
\begin{align}
\mu M&=0&&\; |\Delta_{\text{Re}}|\sim a^2\;& |\Delta_{\text{Im}}|\sim a^{2.3} \nonumber \\
\mu M&=0.1&&\; |\Delta_{\text{Re}}|\sim a^{2.1}\;& |\Delta_{\text{Im}}|\sim a^{2.7} \nonumber \\
\mu M&=0.2&&\; |\Delta_{\text{Re}}|\sim a^{2.5}\;& |\Delta_{\text{Im}}|\sim a^{0.7}
\end{align}
where $|\Delta_{\text{Re}}|$ and $|\Delta_{\text{Im}}|$ are the absolute magnitudes of the percentage error in the real and imaginary parts of the frequencies respectively. Most of the errors appear to be dominated by a component quadratic in spin $a$, indicating that most of the linear in spin dependence has been accounted for in the $L$ expansion. This is not the case for $|\Delta_{\text{Im}}|$ for $\mu M=0.2$, however, indicating that there are more nuanced dependencies on the effective scalar mass $\mu$ and the angular momentum $a$ that are not captured by the first order approximation used in this section. This is, of course, expected given the limitations of applying the expansion method used here to the rotating regime (as discussed above).

\begin{figure}
\caption{Real and imaginary frequency components for the $\ell=m=1,n=0$ mode as a function of $\mu$ for different values of dimensionless angular momentum $a$.}
\label{fig4}
\includegraphics[width=0.5\textwidth]{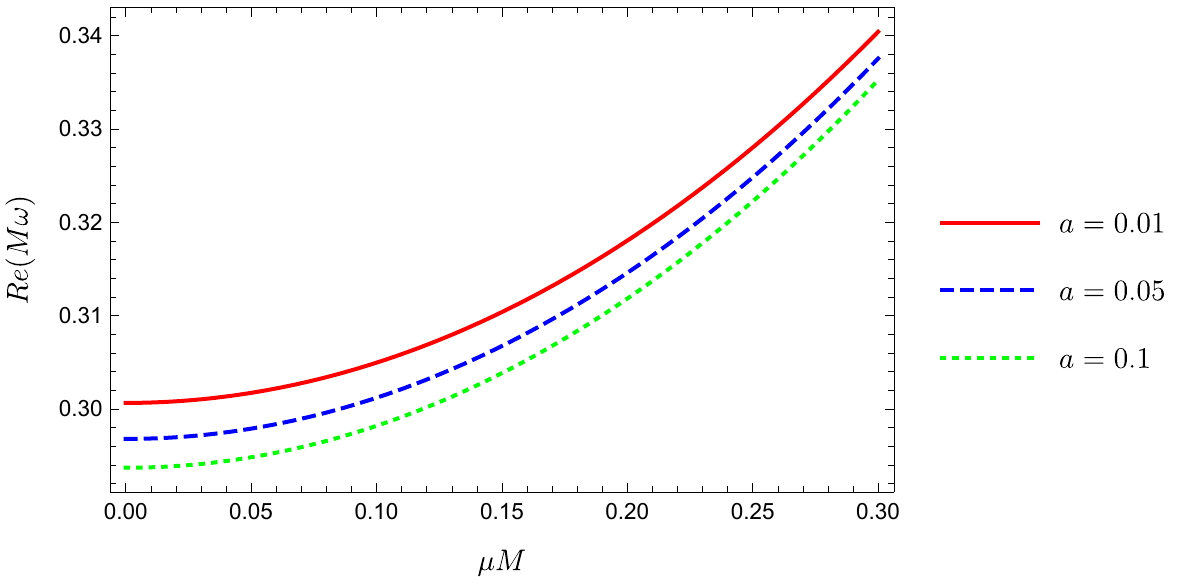}\\
\includegraphics[width=0.5\textwidth]{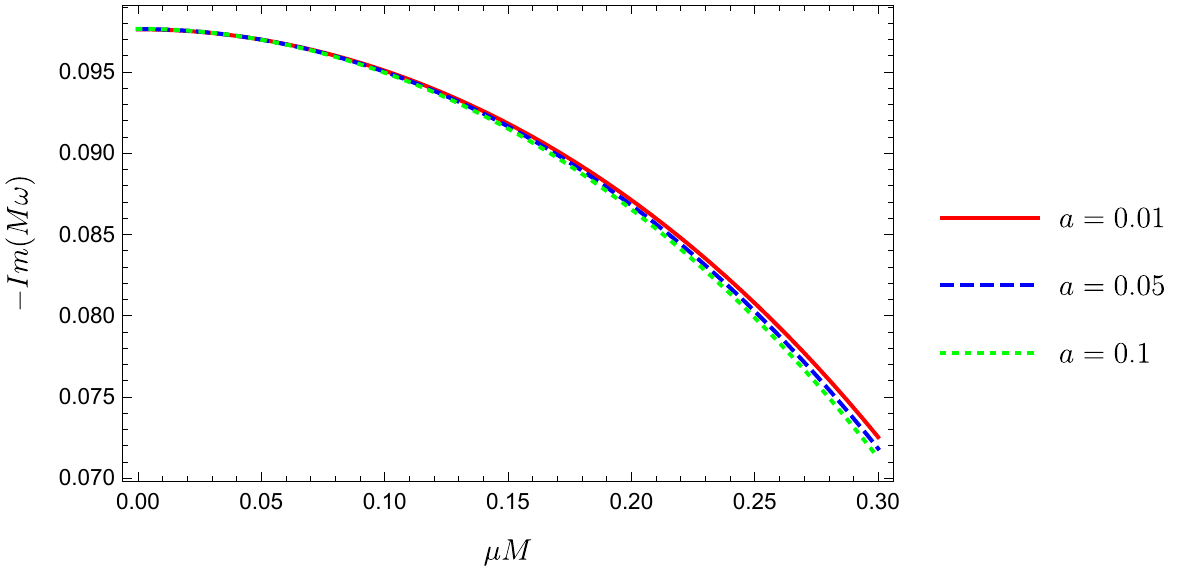}
\end{figure}

In spite of the accuracy limitations discussed above, the expansion provided by eq.~(\ref{expansion2})-(\ref{omegaSR}) provides a useful tool for calculating the scalar QNMs that may appear in the gravitational wave signature of a perturbed, slowly rotating black hole due to the conformal coupling between $\phi$ and the metric (as discussed above). Figure \ref{fig4} shows the real and imaginary components of the QNMs calculated using eq.~(\ref{expansion2})-(\ref{omegaSR}) as a function of $\mu$ for different (small) values of $a$. 

\section{Conclusion}\label{conclusion}

In this paper we have analysed the equations of motion for perturbations to a Schwarzschild black hole in Horndeski gravity, showing that whilst the background is identical to the corresponding solution in GR, the gravitational waves of such a perturbed system can be `contaminated' with non-GR frequencies arising from the spectrum associated with the non-minimally coupled scalar field perturbation. This is due to the conformal coupling between the scalar field and curvature, with the even parity metric perturbations being given in terms of \textit{both} a `Zerilli' function \textit{and} the Horndeski scalar field perturbation (see eq.~(\ref{eqmixing})). This effect was noted in \cite{2018PhRvD..97d4021T}. If there was no conformal coupling then the metric and scalar perturbations would, however, be uncoupled, and thus gravitational waves would not be modified from their GR form (e.g. in ESGB gravity). 

We showed that the non-GR scalar field perturbation obeys a modified massive Klein-Gordon equation characterised by two parameters $\mu$ and $\Gamma$, which are in turn given in terms of the free functions $G_i$ in the Horndeski Lagrangian (see eq.~(\ref{params})). These two parameters will modify the scalar QNM spectrum from the usual case of a massless test field in GR, and thus provide a way to observationally constrain the Horndeski functions $G_i$ through ringdown observations. 

Using the expansion in $L=\ell+1/2$ method developed in \cite{2009CQGra..26v5003D}, we calculate analytical expressions for the QNMs from the scalar spectrum, for arbitrary $\ell$ and overtone index $n$ (provided $\ell>1, \ell>n$) for the subclass of Horndeski theories which satisfy the constraint $c_T=1$ \cite{Lombriser:2015sxa,Lombriser:2016yzn,2017PhRvL.119y1301B,Creminelli:2017sry,Sakstein:2017xjx,Ezquiaga:2017ekz}. For this subclass we find that one of the two free parameters present in the scalar equation of motion, $\Gamma$, vanishes, leaving $\mu$ as an `effective mass' as the only free parameter left characterising the QNM spectrum. QNMs are calculated for varying $\ell$, $n$, and $\mu$, and are shown to be in strong (i.e. sub-percent error) agreement with those calculated numerically in \cite{2006PhRvD..73l4040K}, despite the relatively small number of terms calculated in the analytical expansion. We further extend the analytical expressions derived for perturbations to a Schwarzschild black hole to a slowly rotating (i.e. $|a|<<1$) Kerr black hole, and again show that the QNMs calculated analytically agree well with those frequencies calculated numerically for $a\leq0.2$. 

The analytical expressions for QNMs provided here are, of course, limited in their scope. As mentioned above, the technique developed by \cite{2009CQGra..26v5003D} is valid for $\ell>1, \ell>n$. For static, spherically symmetric black holes we further find that the accuracy of our expressions decreased with increasing scalar mass $\mu$. We believe that this is due to neglected higher order terms in the expansions becoming more relevant with increasing $\mu$, thus further terms in the expansion should be calculated to yield more accurate results at high masses. Furthermore, the expressions given in eq.~({\ref{omegaSR}) for the scalar QNMs of a slowly rotating Kerr black hole suffer from increasing inaccuracies as the black hole spin increases. This is wholly predictable given the slow rotation approximation used, as well as the `massaging' required to apply the technique of \cite{2009CQGra..26v5003D} to the slowly rotating Kerr background (as discussed in Section \ref{secSR}) .

Nonetheless the main result shown in this paper, that even `bald' black holes in Horndeski gravity (i.e. those that are identical to their GR counterparts) can exhibit a modified gravitational wave signal during ringdown which is characterised by just one or two parameters, is a useful observation for attempts on constraining gravity in this new era of gravitational wave astronomy. The resolving of multiple QNMs in a ringdown signal has already been suggested as a way to test the `no-hair theorem' of black holes \cite{Berti:2005ys,Berti:2009kk,Dreyer:2003bv}, however as discussed similar observations can also probe black holes without hair. For example, with a Schwarzschild black hole with unit mass, it is easy to calculate that the $l=2,n=0$ mode \textit{from the scalar spectrum} is approximately $0.496 -0.092i$ with $\mu M=0.2$. The quality factor $Q=Re(\omega)/2|Im(\omega)|$ (a rough measure of the number of oscillations in one $e$-folding time \cite{Berti:2018vdi}) of this mode is $2.69$. This has a comparable damping time and quality factor to the $l=3,n=0$ mode from the standard GR spectrum ($0.599-0.093i$ with $Q=3.23$ \cite{2009CQGra..26v5003D}). It is, therefore, conceivable that with detections of just the first few `least damped' modes in a ringdown signal, one could place bounds on the effective mass $\mu$ of the scalar field.

In addition, the analytical expressions for scalar QNMs found here provide a quick and easy method of studying the effect of varying scalar mass $\mu$ and black hole angular momentum on the numerical values of the complex frequencies. We anticipate that such expressions will be a useful addition to the numerous tools available to those studying QNMs.

A natural extension to this work will be to simply derive more terms in the analytical $L$ expansion to calculate the QNMs focussed on in this paper more accurately. More expansively, one could consider theories with non-zero $G_{4X}$ and derive analytical expressions for QNMs for arbitrary $\Gamma$ so as to encompass further subcategories of Horndeski gravity. In addition, the detectability of these types of non-GR effects in the ringdown signal of a black hole merger remnant need to be explored in further detail. The calculation of QNMs, analytically or otherwise, for scalar tensor theories on a variety of backgrounds (for example a hairy black hole or a more rapidly rotating Kerr black hole) is of course an active and important area of research for the global gravity community.   

\section*{Acknowledgments}
\vspace{-0.2in}
\noindent We are grateful to R. Konoplya for useful discussions and for the sharing of numerical results. OJT was supported by the Science and Technology Facilities Council (STFC) Project Reference 1804725. PGF acknowledges support from STFC, the Beecroft Trust and the European Research Council.

\bibliography{RefModifiedGravity}

\begin{thebibliography}{60}%
\makeatletter
\providecommand \@ifxundefined [1]{%
 \@ifx{#1\undefined}
}%
\providecommand \@ifnum [1]{%
 \ifnum #1\expandafter \@firstoftwo
 \else \expandafter \@secondoftwo
 \fi
}%
\providecommand \@ifx [1]{%
 \ifx #1\expandafter \@firstoftwo
 \else \expandafter \@secondoftwo
 \fi
}%
\providecommand \natexlab [1]{#1}%
\providecommand \enquote  [1]{``#1''}%
\providecommand \bibnamefont  [1]{#1}%
\providecommand \bibfnamefont [1]{#1}%
\providecommand \citenamefont [1]{#1}%
\providecommand \href@noop [0]{\@secondoftwo}%
\providecommand \href [0]{\begingroup \@sanitize@url \@href}%
\providecommand \@href[1]{\@@startlink{#1}\@@href}%
\providecommand \@@href[1]{\endgroup#1\@@endlink}%
\providecommand \@sanitize@url [0]{\catcode `\\12\catcode `\$12\catcode
  `\&12\catcode `\#12\catcode `\^12\catcode `\_12\catcode `\%12\relax}%
\providecommand \@@startlink[1]{}%
\providecommand \@@endlink[0]{}%
\providecommand \url  [0]{\begingroup\@sanitize@url \@url }%
\providecommand \@url [1]{\endgroup\@href {#1}{\urlprefix }}%
\providecommand \urlprefix  [0]{URL }%
\providecommand \Eprint [0]{\href }%
\providecommand \doibase [0]{http://dx.doi.org/}%
\providecommand \selectlanguage [0]{\@gobble}%
\providecommand \bibinfo  [0]{\@secondoftwo}%
\providecommand \bibfield  [0]{\@secondoftwo}%
\providecommand \translation [1]{[#1]}%
\providecommand \BibitemOpen [0]{}%
\providecommand \bibitemStop [0]{}%
\providecommand \bibitemNoStop [0]{.\EOS\space}%
\providecommand \EOS [0]{\spacefactor3000\relax}%
\providecommand \BibitemShut  [1]{\csname bibitem#1\endcsname}%
\let\auto@bib@innerbib\@empty
\bibitem [{\citenamefont {Will}(2014)}]{Will:2014kxa}%
  \BibitemOpen
  \bibfield  {author} {\bibinfo {author} {\bibfnamefont {C.~M.}\ \bibnamefont
  {Will}},\ }\href {\doibase 10.12942/lrr-2014-4} {\bibfield  {journal}
  {\bibinfo  {journal} {Living Rev. Rel.}\ }\textbf {\bibinfo {volume} {17}},\
  \bibinfo {pages} {4} (\bibinfo {year} {2014})},\ \Eprint
  {http://arxiv.org/abs/1403.7377} {arXiv:1403.7377 [gr-qc]} \BibitemShut
  {NoStop}%
\bibitem [{\citenamefont {Clifton}\ \emph {et~al.}(2012)\citenamefont
  {Clifton}, \citenamefont {Ferreira}, \citenamefont {Padilla},\ and\
  \citenamefont {Skordis}}]{Clifton20121}%
  \BibitemOpen
  \bibfield  {author} {\bibinfo {author} {\bibfnamefont {T.}~\bibnamefont
  {Clifton}}, \bibinfo {author} {\bibfnamefont {P.~G.}\ \bibnamefont
  {Ferreira}}, \bibinfo {author} {\bibfnamefont {A.}~\bibnamefont {Padilla}}, \
  and\ \bibinfo {author} {\bibfnamefont {C.}~\bibnamefont {Skordis}},\ }\href
  {\doibase https://doi.org/10.1016/j.physrep.2012.01.001} {\bibfield
  {journal} {\bibinfo  {journal} {Physics Reports}\ }\textbf {\bibinfo {volume}
  {513}},\ \bibinfo {pages} {1 } (\bibinfo {year} {2012})},\ \bibinfo {note}
  {modified Gravity and Cosmology}\BibitemShut {NoStop}%
\bibitem [{\citenamefont {de~Rham}\ and\ \citenamefont
  {Heisenberg}(2011)}]{deRham:2011by}%
  \BibitemOpen
  \bibfield  {author} {\bibinfo {author} {\bibfnamefont {C.}~\bibnamefont
  {de~Rham}}\ and\ \bibinfo {author} {\bibfnamefont {L.}~\bibnamefont
  {Heisenberg}},\ }\href {\doibase 10.1103/PhysRevD.84.043503} {\bibfield
  {journal} {\bibinfo  {journal} {Phys. Rev.}\ }\textbf {\bibinfo {volume}
  {D84}},\ \bibinfo {pages} {043503} (\bibinfo {year} {2011})},\ \Eprint
  {http://arxiv.org/abs/1106.3312} {arXiv:1106.3312 [hep-th]} \BibitemShut
  {NoStop}%
\bibitem [{\citenamefont {Berti}\ \emph {et~al.}(2018)\citenamefont {Berti},
  \citenamefont {Yagi}, \citenamefont {Yang},\ and\ \citenamefont
  {Yunes}}]{Berti:2018vdi}%
  \BibitemOpen
  \bibfield  {author} {\bibinfo {author} {\bibfnamefont {E.}~\bibnamefont
  {Berti}}, \bibinfo {author} {\bibfnamefont {K.}~\bibnamefont {Yagi}},
  \bibinfo {author} {\bibfnamefont {H.}~\bibnamefont {Yang}}, \ and\ \bibinfo
  {author} {\bibfnamefont {N.}~\bibnamefont {Yunes}},\ }\href@noop {} {\
  (\bibinfo {year} {2018})},\ \Eprint {http://arxiv.org/abs/1801.03587}
  {arXiv:1801.03587 [gr-qc]} \BibitemShut {NoStop}%
\bibitem [{\citenamefont {Kokkotas}\ and\ \citenamefont
  {Schmidt}(1999)}]{Kokkotas:1999bd}%
  \BibitemOpen
  \bibfield  {author} {\bibinfo {author} {\bibfnamefont {K.~D.}\ \bibnamefont
  {Kokkotas}}\ and\ \bibinfo {author} {\bibfnamefont {B.~G.}\ \bibnamefont
  {Schmidt}},\ }\href {\doibase 10.12942/lrr-1999-2} {\bibfield  {journal}
  {\bibinfo  {journal} {Living Rev. Rel.}\ }\textbf {\bibinfo {volume} {2}},\
  \bibinfo {pages} {2} (\bibinfo {year} {1999})},\ \Eprint
  {http://arxiv.org/abs/gr-qc/9909058} {arXiv:gr-qc/9909058 [gr-qc]}
  \BibitemShut {NoStop}%
\bibitem [{\citenamefont {Nollert}(1999)}]{0264-9381-16-12-201}%
  \BibitemOpen
  \bibfield  {author} {\bibinfo {author} {\bibfnamefont {H.-P.}\ \bibnamefont
  {Nollert}},\ }\href {http://stacks.iop.org/0264-9381/16/i=12/a=201}
  {\bibfield  {journal} {\bibinfo  {journal} {Classical and Quantum Gravity}\
  }\textbf {\bibinfo {volume} {16}},\ \bibinfo {pages} {R159} (\bibinfo {year}
  {1999})}\BibitemShut {NoStop}%
\bibitem [{\citenamefont {Berti}\ \emph {et~al.}(2009)\citenamefont {Berti},
  \citenamefont {Cardoso},\ and\ \citenamefont {Starinets}}]{Berti:2009kk}%
  \BibitemOpen
  \bibfield  {author} {\bibinfo {author} {\bibfnamefont {E.}~\bibnamefont
  {Berti}}, \bibinfo {author} {\bibfnamefont {V.}~\bibnamefont {Cardoso}}, \
  and\ \bibinfo {author} {\bibfnamefont {A.~O.}\ \bibnamefont {Starinets}},\
  }\href {\doibase 10.1088/0264-9381/26/16/163001} {\bibfield  {journal}
  {\bibinfo  {journal} {Class. Quant. Grav.}\ }\textbf {\bibinfo {volume}
  {26}},\ \bibinfo {pages} {163001} (\bibinfo {year} {2009})},\ \Eprint
  {http://arxiv.org/abs/0905.2975} {arXiv:0905.2975 [gr-qc]} \BibitemShut
  {NoStop}%
\bibitem [{\citenamefont {Berti}\ \emph {et~al.}(2006)\citenamefont {Berti},
  \citenamefont {Cardoso},\ and\ \citenamefont {Will}}]{Berti:2005ys}%
  \BibitemOpen
  \bibfield  {author} {\bibinfo {author} {\bibfnamefont {E.}~\bibnamefont
  {Berti}}, \bibinfo {author} {\bibfnamefont {V.}~\bibnamefont {Cardoso}}, \
  and\ \bibinfo {author} {\bibfnamefont {C.~M.}\ \bibnamefont {Will}},\ }\href
  {\doibase 10.1103/PhysRevD.73.064030} {\bibfield  {journal} {\bibinfo
  {journal} {Phys. Rev.}\ }\textbf {\bibinfo {volume} {D73}},\ \bibinfo {pages}
  {064030} (\bibinfo {year} {2006})},\ \Eprint
  {http://arxiv.org/abs/gr-qc/0512160} {arXiv:gr-qc/0512160 [gr-qc]}
  \BibitemShut {NoStop}%
\bibitem [{\citenamefont {Dreyer}\ \emph {et~al.}(2004)\citenamefont {Dreyer},
  \citenamefont {Kelly}, \citenamefont {Krishnan}, \citenamefont {Finn},
  \citenamefont {Garrison},\ and\ \citenamefont
  {Lopez-Aleman}}]{Dreyer:2003bv}%
  \BibitemOpen
  \bibfield  {author} {\bibinfo {author} {\bibfnamefont {O.}~\bibnamefont
  {Dreyer}}, \bibinfo {author} {\bibfnamefont {B.~J.}\ \bibnamefont {Kelly}},
  \bibinfo {author} {\bibfnamefont {B.}~\bibnamefont {Krishnan}}, \bibinfo
  {author} {\bibfnamefont {L.~S.}\ \bibnamefont {Finn}}, \bibinfo {author}
  {\bibfnamefont {D.}~\bibnamefont {Garrison}}, \ and\ \bibinfo {author}
  {\bibfnamefont {R.}~\bibnamefont {Lopez-Aleman}},\ }\href {\doibase
  10.1088/0264-9381/21/4/003} {\bibfield  {journal} {\bibinfo  {journal}
  {Class. Quant. Grav.}\ }\textbf {\bibinfo {volume} {21}},\ \bibinfo {pages}
  {787} (\bibinfo {year} {2004})},\ \Eprint
  {http://arxiv.org/abs/gr-qc/0309007} {arXiv:gr-qc/0309007 [gr-qc]}
  \BibitemShut {NoStop}%
\bibitem [{\citenamefont {{Abbott}}\ \emph
  {et~al.}(2016{\natexlab{a}})\citenamefont {{Abbott}}, \citenamefont
  {{Abbott}}, \citenamefont {{Abbott}}, \citenamefont {{Abernathy}},
  \citenamefont {{Acernese}}, \citenamefont {{Ackley}}, \citenamefont
  {{Adams}}, \citenamefont {{Adams}}, \citenamefont {{Addesso}}, \citenamefont
  {{Adhikari}},\ and\ \citenamefont {et~al.}}]{2016PhRvL.116f1102A}%
  \BibitemOpen
  \bibfield  {author} {\bibinfo {author} {\bibfnamefont {B.~P.}\ \bibnamefont
  {{Abbott}}}, \bibinfo {author} {\bibfnamefont {R.}~\bibnamefont {{Abbott}}},
  \bibinfo {author} {\bibfnamefont {T.~D.}\ \bibnamefont {{Abbott}}}, \bibinfo
  {author} {\bibfnamefont {M.~R.}\ \bibnamefont {{Abernathy}}}, \bibinfo
  {author} {\bibfnamefont {F.}~\bibnamefont {{Acernese}}}, \bibinfo {author}
  {\bibfnamefont {K.}~\bibnamefont {{Ackley}}}, \bibinfo {author}
  {\bibfnamefont {C.}~\bibnamefont {{Adams}}}, \bibinfo {author} {\bibfnamefont
  {T.}~\bibnamefont {{Adams}}}, \bibinfo {author} {\bibfnamefont
  {P.}~\bibnamefont {{Addesso}}}, \bibinfo {author} {\bibfnamefont {R.~X.}\
  \bibnamefont {{Adhikari}}}, \ and\ \bibinfo {author} {\bibnamefont
  {et~al.}},\ }\href {\doibase 10.1103/PhysRevLett.116.061102} {\bibfield
  {journal} {\bibinfo  {journal} {Physical Review Letters}\ }\textbf {\bibinfo
  {volume} {116}},\ \bibinfo {eid} {061102} (\bibinfo {year}
  {2016}{\natexlab{a}})},\ \Eprint {http://arxiv.org/abs/1602.03837}
  {arXiv:1602.03837 [gr-qc]} \BibitemShut {NoStop}%
\bibitem [{\citenamefont {{Abbott}}\ \emph
  {et~al.}(2016{\natexlab{b}})\citenamefont {{Abbott}}, \citenamefont
  {{Abbott}}, \citenamefont {{Abbott}}, \citenamefont {{Abernathy}},
  \citenamefont {{Acernese}}, \citenamefont {{Ackley}}, \citenamefont
  {{Adams}}, \citenamefont {{Adams}}, \citenamefont {{Addesso}}, \citenamefont
  {{Adhikari}},\ and\ \citenamefont {et~al.}}]{2016PhRvL.116x1103A}%
  \BibitemOpen
  \bibfield  {author} {\bibinfo {author} {\bibfnamefont {B.~P.}\ \bibnamefont
  {{Abbott}}}, \bibinfo {author} {\bibfnamefont {R.}~\bibnamefont {{Abbott}}},
  \bibinfo {author} {\bibfnamefont {T.~D.}\ \bibnamefont {{Abbott}}}, \bibinfo
  {author} {\bibfnamefont {M.~R.}\ \bibnamefont {{Abernathy}}}, \bibinfo
  {author} {\bibfnamefont {F.}~\bibnamefont {{Acernese}}}, \bibinfo {author}
  {\bibfnamefont {K.}~\bibnamefont {{Ackley}}}, \bibinfo {author}
  {\bibfnamefont {C.}~\bibnamefont {{Adams}}}, \bibinfo {author} {\bibfnamefont
  {T.}~\bibnamefont {{Adams}}}, \bibinfo {author} {\bibfnamefont
  {P.}~\bibnamefont {{Addesso}}}, \bibinfo {author} {\bibfnamefont {R.~X.}\
  \bibnamefont {{Adhikari}}}, \ and\ \bibinfo {author} {\bibnamefont
  {et~al.}},\ }\href {\doibase 10.1103/PhysRevLett.116.241103} {\bibfield
  {journal} {\bibinfo  {journal} {Physical Review Letters}\ }\textbf {\bibinfo
  {volume} {116}},\ \bibinfo {eid} {241103} (\bibinfo {year}
  {2016}{\natexlab{b}})},\ \Eprint {http://arxiv.org/abs/1606.04855}
  {arXiv:1606.04855 [gr-qc]} \BibitemShut {NoStop}%
\bibitem [{\citenamefont {{Abbott}}\ \emph {et~al.}(2017)\citenamefont
  {{Abbott}}, \citenamefont {{Abbott}}, \citenamefont {{Abbott}}, \citenamefont
  {{Acernese}}, \citenamefont {{Ackley}}, \citenamefont {{Adams}},
  \citenamefont {{Adams}}, \citenamefont {{Addesso}}, \citenamefont
  {{Adhikari}}, \citenamefont {{Adya}},\ and\ \citenamefont
  {et~al.}}]{2017PhRvL.118v1101A}%
  \BibitemOpen
  \bibfield  {author} {\bibinfo {author} {\bibfnamefont {B.~P.}\ \bibnamefont
  {{Abbott}}}, \bibinfo {author} {\bibfnamefont {R.}~\bibnamefont {{Abbott}}},
  \bibinfo {author} {\bibfnamefont {T.~D.}\ \bibnamefont {{Abbott}}}, \bibinfo
  {author} {\bibfnamefont {F.}~\bibnamefont {{Acernese}}}, \bibinfo {author}
  {\bibfnamefont {K.}~\bibnamefont {{Ackley}}}, \bibinfo {author}
  {\bibfnamefont {C.}~\bibnamefont {{Adams}}}, \bibinfo {author} {\bibfnamefont
  {T.}~\bibnamefont {{Adams}}}, \bibinfo {author} {\bibfnamefont
  {P.}~\bibnamefont {{Addesso}}}, \bibinfo {author} {\bibfnamefont {R.~X.}\
  \bibnamefont {{Adhikari}}}, \bibinfo {author} {\bibfnamefont {V.~B.}\
  \bibnamefont {{Adya}}}, \ and\ \bibinfo {author} {\bibnamefont {et~al.}},\
  }\href {\doibase 10.1103/PhysRevLett.118.221101} {\bibfield  {journal}
  {\bibinfo  {journal} {Physical Review Letters}\ }\textbf {\bibinfo {volume}
  {118}},\ \bibinfo {eid} {221101} (\bibinfo {year} {2017})},\ \Eprint
  {http://arxiv.org/abs/1706.01812} {arXiv:1706.01812 [gr-qc]} \BibitemShut
  {NoStop}%
\bibitem [{\citenamefont {Abbott}\ \emph {et~al.}(2017)\citenamefont {Abbott}
  \emph {et~al.}}]{Abbott:2017oio}%
  \BibitemOpen
  \bibfield  {author} {\bibinfo {author} {\bibfnamefont {B.~P.}\ \bibnamefont
  {Abbott}} \emph {et~al.} (\bibinfo {collaboration} {Virgo, LIGO
  Scientific}),\ }\href {\doibase 10.1103/PhysRevLett.119.141101} {\bibfield
  {journal} {\bibinfo  {journal} {Phys. Rev. Lett.}\ }\textbf {\bibinfo
  {volume} {119}},\ \bibinfo {pages} {141101} (\bibinfo {year} {2017})},\
  \Eprint {http://arxiv.org/abs/1709.09660} {arXiv:1709.09660 [gr-qc]}
  \BibitemShut {NoStop}%
\bibitem [{\citenamefont {et.
  al.}(2017{\natexlab{a}})}]{PhysRevLett.119.161101}%
  \BibitemOpen
  \bibfield  {author} {\bibinfo {author} {\bibfnamefont {B.~P.~A.}\
  \bibnamefont {et. al.}} (\bibinfo {collaboration} {LIGO Scientific
  Collaboration and Virgo Collaboration}),\ }\href {\doibase
  10.1103/PhysRevLett.119.161101} {\bibfield  {journal} {\bibinfo  {journal}
  {Phys. Rev. Lett.}\ }\textbf {\bibinfo {volume} {119}},\ \bibinfo {pages}
  {161101} (\bibinfo {year} {2017}{\natexlab{a}})}\BibitemShut {NoStop}%
\bibitem [{\citenamefont {Horndeski}(1974)}]{Horndeski:1974wa}%
  \BibitemOpen
  \bibfield  {author} {\bibinfo {author} {\bibfnamefont {G.~W.}\ \bibnamefont
  {Horndeski}},\ }\href {\doibase 10.1007/BF01807638} {\bibfield  {journal}
  {\bibinfo  {journal} {Int. J. Theor. Phys.}\ }\textbf {\bibinfo {volume}
  {10}},\ \bibinfo {pages} {363} (\bibinfo {year} {1974})}\BibitemShut
  {NoStop}%
\bibitem [{\citenamefont {{Dolan}}\ and\ \citenamefont
  {{Ottewill}}(2009)}]{2009CQGra..26v5003D}%
  \BibitemOpen
  \bibfield  {author} {\bibinfo {author} {\bibfnamefont {S.~R.}\ \bibnamefont
  {{Dolan}}}\ and\ \bibinfo {author} {\bibfnamefont {A.~C.}\ \bibnamefont
  {{Ottewill}}},\ }\href {\doibase 10.1088/0264-9381/26/22/225003} {\bibfield
  {journal} {\bibinfo  {journal} {Classical and Quantum Gravity}\ }\textbf
  {\bibinfo {volume} {26}},\ \bibinfo {eid} {225003} (\bibinfo {year}
  {2009})},\ \Eprint {http://arxiv.org/abs/0908.0329} {arXiv:0908.0329 [gr-qc]}
  \BibitemShut {NoStop}%
\bibitem [{\citenamefont {Konoplya}\ and\ \citenamefont
  {Zhidenko}(2005)}]{Konoplya:2004wg}%
  \BibitemOpen
  \bibfield  {author} {\bibinfo {author} {\bibfnamefont {R.~A.}\ \bibnamefont
  {Konoplya}}\ and\ \bibinfo {author} {\bibfnamefont {A.~V.}\ \bibnamefont
  {Zhidenko}},\ }\href {\doibase 10.1016/j.physletb.2005.01.078} {\bibfield
  {journal} {\bibinfo  {journal} {Phys. Lett.}\ }\textbf {\bibinfo {volume}
  {B609}},\ \bibinfo {pages} {377} (\bibinfo {year} {2005})},\ \Eprint
  {http://arxiv.org/abs/gr-qc/0411059} {arXiv:gr-qc/0411059 [gr-qc]}
  \BibitemShut {NoStop}%
\bibitem [{\citenamefont {{Konoplya}}\ and\ \citenamefont
  {{Zhidenko}}(2006)}]{2006PhRvD..73l4040K}%
  \BibitemOpen
  \bibfield  {author} {\bibinfo {author} {\bibfnamefont {R.~A.}\ \bibnamefont
  {{Konoplya}}}\ and\ \bibinfo {author} {\bibfnamefont {A.~V.}\ \bibnamefont
  {{Zhidenko}}},\ }\href {\doibase 10.1103/PhysRevD.73.124040} {\bibfield
  {journal} {\bibinfo  {journal} {\prd}\ }\textbf {\bibinfo {volume} {73}},\
  \bibinfo {eid} {124040} (\bibinfo {year} {2006})},\ \Eprint
  {http://arxiv.org/abs/gr-qc/0605013} {gr-qc/0605013} \BibitemShut {NoStop}%
\bibitem [{\citenamefont {Zumalacarregui}\ and\ \citenamefont
  {Garcia-Bellido}(2014)}]{Zumalacarregui:2013pma}%
  \BibitemOpen
  \bibfield  {author} {\bibinfo {author} {\bibfnamefont {M.}~\bibnamefont
  {Zumalacarregui}}\ and\ \bibinfo {author} {\bibfnamefont {J.}~\bibnamefont
  {Garcia-Bellido}},\ }\href {\doibase 10.1103/PhysRevD.89.064046} {\bibfield
  {journal} {\bibinfo  {journal} {Phys. Rev.}\ }\textbf {\bibinfo {volume}
  {D89}},\ \bibinfo {pages} {064046} (\bibinfo {year} {2014})},\ \Eprint
  {http://arxiv.org/abs/1308.4685} {arXiv:1308.4685 [gr-qc]} \BibitemShut
  {NoStop}%
\bibitem [{\citenamefont {Gleyzes}\ \emph
  {et~al.}(2015{\natexlab{a}})\citenamefont {Gleyzes}, \citenamefont
  {Langlois}, \citenamefont {Piazza},\ and\ \citenamefont
  {Vernizzi}}]{Gleyzes:2014qga}%
  \BibitemOpen
  \bibfield  {author} {\bibinfo {author} {\bibfnamefont {J.}~\bibnamefont
  {Gleyzes}}, \bibinfo {author} {\bibfnamefont {D.}~\bibnamefont {Langlois}},
  \bibinfo {author} {\bibfnamefont {F.}~\bibnamefont {Piazza}}, \ and\ \bibinfo
  {author} {\bibfnamefont {F.}~\bibnamefont {Vernizzi}},\ }\href {\doibase
  10.1088/1475-7516/2015/02/018} {\bibfield  {journal} {\bibinfo  {journal}
  {JCAP}\ }\textbf {\bibinfo {volume} {1502}},\ \bibinfo {pages} {018}
  (\bibinfo {year} {2015}{\natexlab{a}})},\ \Eprint
  {http://arxiv.org/abs/1408.1952} {arXiv:1408.1952 [astro-ph.CO]} \BibitemShut
  {NoStop}%
\bibitem [{\citenamefont {Gleyzes}\ \emph
  {et~al.}(2015{\natexlab{b}})\citenamefont {Gleyzes}, \citenamefont
  {Langlois}, \citenamefont {Piazza},\ and\ \citenamefont
  {Vernizzi}}]{Gleyzes:2014dya}%
  \BibitemOpen
  \bibfield  {author} {\bibinfo {author} {\bibfnamefont {J.}~\bibnamefont
  {Gleyzes}}, \bibinfo {author} {\bibfnamefont {D.}~\bibnamefont {Langlois}},
  \bibinfo {author} {\bibfnamefont {F.}~\bibnamefont {Piazza}}, \ and\ \bibinfo
  {author} {\bibfnamefont {F.}~\bibnamefont {Vernizzi}},\ }\href {\doibase
  10.1103/PhysRevLett.114.211101} {\bibfield  {journal} {\bibinfo  {journal}
  {Phys. Rev. Lett.}\ }\textbf {\bibinfo {volume} {114}},\ \bibinfo {pages}
  {211101} (\bibinfo {year} {2015}{\natexlab{b}})},\ \Eprint
  {http://arxiv.org/abs/1404.6495} {arXiv:1404.6495 [hep-th]} \BibitemShut
  {NoStop}%
\bibitem [{\citenamefont {Ben~Achour}\ \emph {et~al.}(2016)\citenamefont
  {Ben~Achour}, \citenamefont {Langlois},\ and\ \citenamefont
  {Noui}}]{Achour:2016rkg}%
  \BibitemOpen
  \bibfield  {author} {\bibinfo {author} {\bibfnamefont {J.}~\bibnamefont
  {Ben~Achour}}, \bibinfo {author} {\bibfnamefont {D.}~\bibnamefont
  {Langlois}}, \ and\ \bibinfo {author} {\bibfnamefont {K.}~\bibnamefont
  {Noui}},\ }\href {\doibase 10.1103/PhysRevD.93.124005} {\bibfield  {journal}
  {\bibinfo  {journal} {Phys. Rev.}\ }\textbf {\bibinfo {volume} {D93}},\
  \bibinfo {pages} {124005} (\bibinfo {year} {2016})},\ \Eprint
  {http://arxiv.org/abs/1602.08398} {arXiv:1602.08398 [gr-qc]} \BibitemShut
  {NoStop}%
\bibitem [{\citenamefont {Hui}\ and\ \citenamefont
  {Nicolis}(2013)}]{Hui:2012qt}%
  \BibitemOpen
  \bibfield  {author} {\bibinfo {author} {\bibfnamefont {L.}~\bibnamefont
  {Hui}}\ and\ \bibinfo {author} {\bibfnamefont {A.}~\bibnamefont {Nicolis}},\
  }\href {\doibase 10.1103/PhysRevLett.110.241104} {\bibfield  {journal}
  {\bibinfo  {journal} {Phys. Rev. Lett.}\ }\textbf {\bibinfo {volume} {110}},\
  \bibinfo {pages} {241104} (\bibinfo {year} {2013})},\ \Eprint
  {http://arxiv.org/abs/1202.1296} {arXiv:1202.1296 [hep-th]} \BibitemShut
  {NoStop}%
\bibitem [{\citenamefont {Sotiriou}(2015)}]{Sotiriou:2015pka}%
  \BibitemOpen
  \bibfield  {author} {\bibinfo {author} {\bibfnamefont {T.~P.}\ \bibnamefont
  {Sotiriou}},\ }\href {\doibase 10.1088/0264-9381/32/21/214002} {\bibfield
  {journal} {\bibinfo  {journal} {Class. Quant. Grav.}\ }\textbf {\bibinfo
  {volume} {32}},\ \bibinfo {pages} {214002} (\bibinfo {year} {2015})},\
  \Eprint {http://arxiv.org/abs/1505.00248} {arXiv:1505.00248 [gr-qc]}
  \BibitemShut {NoStop}%
\bibitem [{\citenamefont {Maselli}\ \emph {et~al.}(2015)\citenamefont
  {Maselli}, \citenamefont {Silva}, \citenamefont {Minamitsuji},\ and\
  \citenamefont {Berti}}]{Maselli:2015yva}%
  \BibitemOpen
  \bibfield  {author} {\bibinfo {author} {\bibfnamefont {A.}~\bibnamefont
  {Maselli}}, \bibinfo {author} {\bibfnamefont {H.~O.}\ \bibnamefont {Silva}},
  \bibinfo {author} {\bibfnamefont {M.}~\bibnamefont {Minamitsuji}}, \ and\
  \bibinfo {author} {\bibfnamefont {E.}~\bibnamefont {Berti}},\ }\href
  {\doibase 10.1103/PhysRevD.92.104049} {\bibfield  {journal} {\bibinfo
  {journal} {Phys. Rev.}\ }\textbf {\bibinfo {volume} {D92}},\ \bibinfo {pages}
  {104049} (\bibinfo {year} {2015})},\ \Eprint
  {http://arxiv.org/abs/1508.03044} {arXiv:1508.03044 [gr-qc]} \BibitemShut
  {NoStop}%
\bibitem [{\citenamefont {Brans}\ and\ \citenamefont
  {Dicke}(1961)}]{PhysRev.124.925}%
  \BibitemOpen
  \bibfield  {author} {\bibinfo {author} {\bibfnamefont {C.}~\bibnamefont
  {Brans}}\ and\ \bibinfo {author} {\bibfnamefont {R.~H.}\ \bibnamefont
  {Dicke}},\ }\href {\doibase 10.1103/PhysRev.124.925} {\bibfield  {journal}
  {\bibinfo  {journal} {Phys. Rev.}\ }\textbf {\bibinfo {volume} {124}},\
  \bibinfo {pages} {925} (\bibinfo {year} {1961})}\BibitemShut {NoStop}%
\bibitem [{\citenamefont {{Graham}}\ and\ \citenamefont
  {{Jha}}(2014)}]{2014PhRvD..89h4056G}%
  \BibitemOpen
  \bibfield  {author} {\bibinfo {author} {\bibfnamefont {A.~A.~H.}\
  \bibnamefont {{Graham}}}\ and\ \bibinfo {author} {\bibfnamefont
  {R.}~\bibnamefont {{Jha}}},\ }\href {\doibase 10.1103/PhysRevD.89.084056}
  {\bibfield  {journal} {\bibinfo  {journal} {\prd}\ }\textbf {\bibinfo
  {volume} {89}},\ \bibinfo {eid} {084056} (\bibinfo {year} {2014})},\ \Eprint
  {http://arxiv.org/abs/1401.8203} {arXiv:1401.8203 [gr-qc]} \BibitemShut
  {NoStop}%
\bibitem [{\citenamefont {Tattersall}\ \emph {et~al.}(2018)\citenamefont
  {Tattersall}, \citenamefont {Ferreira},\ and\ \citenamefont
  {Lagos}}]{PhysRevD.97.084005}%
  \BibitemOpen
  \bibfield  {author} {\bibinfo {author} {\bibfnamefont {O.~J.}\ \bibnamefont
  {Tattersall}}, \bibinfo {author} {\bibfnamefont {P.~G.}\ \bibnamefont
  {Ferreira}}, \ and\ \bibinfo {author} {\bibfnamefont {M.}~\bibnamefont
  {Lagos}},\ }\href {\doibase 10.1103/PhysRevD.97.084005} {\bibfield  {journal}
  {\bibinfo  {journal} {Phys. Rev. D}\ }\textbf {\bibinfo {volume} {97}},\
  \bibinfo {pages} {084005} (\bibinfo {year} {2018})}\BibitemShut {NoStop}%
\bibitem [{\citenamefont {Motohashi}\ and\ \citenamefont
  {Minamitsuji}(2018)}]{Motohashi:2018wdq}%
  \BibitemOpen
  \bibfield  {author} {\bibinfo {author} {\bibfnamefont {H.}~\bibnamefont
  {Motohashi}}\ and\ \bibinfo {author} {\bibfnamefont {M.}~\bibnamefont
  {Minamitsuji}},\ }\href@noop {} {\  (\bibinfo {year} {2018})},\ \Eprint
  {http://arxiv.org/abs/1804.01731} {arXiv:1804.01731 [gr-qc]} \BibitemShut
  {NoStop}%
\bibitem [{\citenamefont {{Tattersall}}\ \emph {et~al.}(2018)\citenamefont
  {{Tattersall}}, \citenamefont {{Ferreira}},\ and\ \citenamefont
  {{Lagos}}}]{2018PhRvD..97d4021T}%
  \BibitemOpen
  \bibfield  {author} {\bibinfo {author} {\bibfnamefont {O.~J.}\ \bibnamefont
  {{Tattersall}}}, \bibinfo {author} {\bibfnamefont {P.~G.}\ \bibnamefont
  {{Ferreira}}}, \ and\ \bibinfo {author} {\bibfnamefont {M.}~\bibnamefont
  {{Lagos}}},\ }\href {\doibase 10.1103/PhysRevD.97.044021} {\bibfield
  {journal} {\bibinfo  {journal} {\prd}\ }\textbf {\bibinfo {volume} {97}},\
  \bibinfo {eid} {044021} (\bibinfo {year} {2018})},\ \Eprint
  {http://arxiv.org/abs/1711.01992} {arXiv:1711.01992 [gr-qc]} \BibitemShut
  {NoStop}%
\bibitem [{\citenamefont {Regge}\ and\ \citenamefont
  {Wheeler}(1957)}]{Regge:1957td}%
  \BibitemOpen
  \bibfield  {author} {\bibinfo {author} {\bibfnamefont {T.}~\bibnamefont
  {Regge}}\ and\ \bibinfo {author} {\bibfnamefont {J.~A.}\ \bibnamefont
  {Wheeler}},\ }\href {\doibase 10.1103/PhysRev.108.1063} {\bibfield  {journal}
  {\bibinfo  {journal} {Phys. Rev.}\ }\textbf {\bibinfo {volume} {108}},\
  \bibinfo {pages} {1063} (\bibinfo {year} {1957})}\BibitemShut {NoStop}%
\bibitem [{\citenamefont {Rezzolla}(2003)}]{Rezzolla:2003ua}%
  \BibitemOpen
  \bibfield  {author} {\bibinfo {author} {\bibfnamefont {L.}~\bibnamefont
  {Rezzolla}},\ }\bibfield  {booktitle} {\emph {\bibinfo {booktitle}
  {{Astroparticle physics and cosmology. Proceedings: Summer School, Trieste,
  Italy, Jun 17-Jul 5 2002}}},\ }\href@noop {} {\bibfield  {journal} {\bibinfo
  {journal} {ICTP Lect. Notes Ser.}\ }\textbf {\bibinfo {volume} {14}},\
  \bibinfo {pages} {255} (\bibinfo {year} {2003})},\ \Eprint
  {http://arxiv.org/abs/gr-qc/0302025} {arXiv:gr-qc/0302025 [gr-qc]}
  \BibitemShut {NoStop}%
\bibitem [{\citenamefont {Martel}\ and\ \citenamefont
  {Poisson}(2005)}]{Martel:2005ir}%
  \BibitemOpen
  \bibfield  {author} {\bibinfo {author} {\bibfnamefont {K.}~\bibnamefont
  {Martel}}\ and\ \bibinfo {author} {\bibfnamefont {E.}~\bibnamefont
  {Poisson}},\ }\href {\doibase 10.1103/PhysRevD.71.104003} {\bibfield
  {journal} {\bibinfo  {journal} {Phys. Rev.}\ }\textbf {\bibinfo {volume}
  {D71}},\ \bibinfo {pages} {104003} (\bibinfo {year} {2005})},\ \Eprint
  {http://arxiv.org/abs/gr-qc/0502028} {arXiv:gr-qc/0502028 [gr-qc]}
  \BibitemShut {NoStop}%
\bibitem [{\citenamefont {Ripley}\ and\ \citenamefont
  {Yagi}(2018)}]{Ripley:2017kqg}%
  \BibitemOpen
  \bibfield  {author} {\bibinfo {author} {\bibfnamefont {J.~L.}\ \bibnamefont
  {Ripley}}\ and\ \bibinfo {author} {\bibfnamefont {K.}~\bibnamefont {Yagi}},\
  }\href {\doibase 10.1103/PhysRevD.97.024009} {\bibfield  {journal} {\bibinfo
  {journal} {Phys. Rev.}\ }\textbf {\bibinfo {volume} {D97}},\ \bibinfo {pages}
  {024009} (\bibinfo {year} {2018})},\ \Eprint
  {http://arxiv.org/abs/1705.03068} {arXiv:1705.03068 [gr-qc]} \BibitemShut
  {NoStop}%
\bibitem [{\citenamefont {Kobayashi}\ \emph {et~al.}(2012)\citenamefont
  {Kobayashi}, \citenamefont {Motohashi},\ and\ \citenamefont
  {Suyama}}]{Kobayashi:2012kh}%
  \BibitemOpen
  \bibfield  {author} {\bibinfo {author} {\bibfnamefont {T.}~\bibnamefont
  {Kobayashi}}, \bibinfo {author} {\bibfnamefont {H.}~\bibnamefont
  {Motohashi}}, \ and\ \bibinfo {author} {\bibfnamefont {T.}~\bibnamefont
  {Suyama}},\ }\href {\doibase 10.1103/PhysRevD.85.084025} {\bibfield
  {journal} {\bibinfo  {journal} {Phys. Rev.}\ }\textbf {\bibinfo {volume}
  {D85}},\ \bibinfo {pages} {084025} (\bibinfo {year} {2012})},\ \Eprint
  {http://arxiv.org/abs/1202.4893} {arXiv:1202.4893 [gr-qc]} \BibitemShut
  {NoStop}%
\bibitem [{\citenamefont {Kobayashi}\ \emph {et~al.}(2014)\citenamefont
  {Kobayashi}, \citenamefont {Motohashi},\ and\ \citenamefont
  {Suyama}}]{Kobayashi:2014wsa}%
  \BibitemOpen
  \bibfield  {author} {\bibinfo {author} {\bibfnamefont {T.}~\bibnamefont
  {Kobayashi}}, \bibinfo {author} {\bibfnamefont {H.}~\bibnamefont
  {Motohashi}}, \ and\ \bibinfo {author} {\bibfnamefont {T.}~\bibnamefont
  {Suyama}},\ }\href {\doibase 10.1103/PhysRevD.89.084042} {\bibfield
  {journal} {\bibinfo  {journal} {Phys. Rev.}\ }\textbf {\bibinfo {volume}
  {D89}},\ \bibinfo {pages} {084042} (\bibinfo {year} {2014})},\ \Eprint
  {http://arxiv.org/abs/1402.6740} {arXiv:1402.6740 [gr-qc]} \BibitemShut
  {NoStop}%
\bibitem [{\citenamefont {{Chandrasekhar}}\ and\ \citenamefont
  {{Detweiler}}(1975)}]{1975RSPSA.344..441C}%
  \BibitemOpen
  \bibfield  {author} {\bibinfo {author} {\bibfnamefont {S.}~\bibnamefont
  {{Chandrasekhar}}}\ and\ \bibinfo {author} {\bibfnamefont {S.}~\bibnamefont
  {{Detweiler}}},\ }\href {\doibase 10.1098/rspa.1975.0112} {\bibfield
  {journal} {\bibinfo  {journal} {Proceedings of the Royal Society of London
  Series A}\ }\textbf {\bibinfo {volume} {344}},\ \bibinfo {pages} {441}
  (\bibinfo {year} {1975})}\BibitemShut {NoStop}%
\bibitem [{\citenamefont {Zerilli}(1970)}]{Zerilli:1970se}%
  \BibitemOpen
  \bibfield  {author} {\bibinfo {author} {\bibfnamefont {F.~J.}\ \bibnamefont
  {Zerilli}},\ }\href {\doibase 10.1103/PhysRevLett.24.737} {\bibfield
  {journal} {\bibinfo  {journal} {Phys. Rev. Lett.}\ }\textbf {\bibinfo
  {volume} {24}},\ \bibinfo {pages} {737} (\bibinfo {year} {1970})}\BibitemShut
  {NoStop}%
\bibitem [{\citenamefont {Kwon}\ \emph {et~al.}(1986)\citenamefont {Kwon},
  \citenamefont {Kim}, \citenamefont {Myung}, \citenamefont {Cho},\ and\
  \citenamefont {Park}}]{PhysRevD.34.333}%
  \BibitemOpen
  \bibfield  {author} {\bibinfo {author} {\bibfnamefont {O.~J.}\ \bibnamefont
  {Kwon}}, \bibinfo {author} {\bibfnamefont {Y.~D.}\ \bibnamefont {Kim}},
  \bibinfo {author} {\bibfnamefont {Y.~S.}\ \bibnamefont {Myung}}, \bibinfo
  {author} {\bibfnamefont {B.~H.}\ \bibnamefont {Cho}}, \ and\ \bibinfo
  {author} {\bibfnamefont {Y.~J.}\ \bibnamefont {Park}},\ }\href {\doibase
  10.1103/PhysRevD.34.333} {\bibfield  {journal} {\bibinfo  {journal} {Phys.
  Rev. D}\ }\textbf {\bibinfo {volume} {34}},\ \bibinfo {pages} {333} (\bibinfo
  {year} {1986})}\BibitemShut {NoStop}%
\bibitem [{\citenamefont {Silva}\ \emph {et~al.}(2018)\citenamefont {Silva},
  \citenamefont {Sakstein}, \citenamefont {Gualtieri}, \citenamefont
  {Sotiriou},\ and\ \citenamefont {Berti}}]{Silva:2017uqg}%
  \BibitemOpen
  \bibfield  {author} {\bibinfo {author} {\bibfnamefont {H.~O.}\ \bibnamefont
  {Silva}}, \bibinfo {author} {\bibfnamefont {J.}~\bibnamefont {Sakstein}},
  \bibinfo {author} {\bibfnamefont {L.}~\bibnamefont {Gualtieri}}, \bibinfo
  {author} {\bibfnamefont {T.~P.}\ \bibnamefont {Sotiriou}}, \ and\ \bibinfo
  {author} {\bibfnamefont {E.}~\bibnamefont {Berti}},\ }\href {\doibase
  10.1103/PhysRevLett.120.131104} {\bibfield  {journal} {\bibinfo  {journal}
  {Phys. Rev. Lett.}\ }\textbf {\bibinfo {volume} {120}},\ \bibinfo {pages}
  {131104} (\bibinfo {year} {2018})},\ \Eprint
  {http://arxiv.org/abs/1711.02080} {arXiv:1711.02080 [gr-qc]} \BibitemShut
  {NoStop}%
\bibitem [{\citenamefont {Gong}\ \emph {et~al.}(2017)\citenamefont {Gong},
  \citenamefont {Papantonopoulos},\ and\ \citenamefont {Yi}}]{Gong:2017kim}%
  \BibitemOpen
  \bibfield  {author} {\bibinfo {author} {\bibfnamefont {Y.}~\bibnamefont
  {Gong}}, \bibinfo {author} {\bibfnamefont {E.}~\bibnamefont
  {Papantonopoulos}}, \ and\ \bibinfo {author} {\bibfnamefont {Z.}~\bibnamefont
  {Yi}},\ }\href@noop {} {\  (\bibinfo {year} {2017})},\ \Eprint
  {http://arxiv.org/abs/1711.04102} {arXiv:1711.04102 [gr-qc]} \BibitemShut
  {NoStop}%
\bibitem [{\citenamefont {et. al.}(2017{\natexlab{b}})}]{2041-8205-848-2-L12}%
  \BibitemOpen
  \bibfield  {author} {\bibinfo {author} {\bibfnamefont {B.~P.~A.}\
  \bibnamefont {et. al.}},\ }\href
  {http://stacks.iop.org/2041-8205/848/i=2/a=L12} {\bibfield  {journal}
  {\bibinfo  {journal} {The Astrophysical Journal Letters}\ }\textbf {\bibinfo
  {volume} {848}},\ \bibinfo {pages} {L12} (\bibinfo {year}
  {2017}{\natexlab{b}})}\BibitemShut {NoStop}%
\bibitem [{\citenamefont {et. al.}(2017{\natexlab{c}})}]{2041-8205-848-2-L13}%
  \BibitemOpen
  \bibfield  {author} {\bibinfo {author} {\bibfnamefont {B.~P.~A.}\
  \bibnamefont {et. al.}},\ }\href
  {http://stacks.iop.org/2041-8205/848/i=2/a=L13} {\bibfield  {journal}
  {\bibinfo  {journal} {The Astrophysical Journal Letters}\ }\textbf {\bibinfo
  {volume} {848}},\ \bibinfo {pages} {L13} (\bibinfo {year}
  {2017}{\natexlab{c}})}\BibitemShut {NoStop}%
\bibitem [{\citenamefont {et. al.}(2017{\natexlab{d}})}]{2041-8205-848-2-L14}%
  \BibitemOpen
  \bibfield  {author} {\bibinfo {author} {\bibfnamefont {A.~G.}\ \bibnamefont
  {et. al.}},\ }\href {http://stacks.iop.org/2041-8205/848/i=2/a=L14}
  {\bibfield  {journal} {\bibinfo  {journal} {The Astrophysical Journal
  Letters}\ }\textbf {\bibinfo {volume} {848}},\ \bibinfo {pages} {L14}
  (\bibinfo {year} {2017}{\natexlab{d}})}\BibitemShut {NoStop}%
\bibitem [{\citenamefont {et. al.}(2017{\natexlab{e}})}]{2041-8205-848-2-L15}%
  \BibitemOpen
  \bibfield  {author} {\bibinfo {author} {\bibfnamefont {V.~S.}\ \bibnamefont
  {et. al.}},\ }\href {http://stacks.iop.org/2041-8205/848/i=2/a=L15}
  {\bibfield  {journal} {\bibinfo  {journal} {The Astrophysical Journal
  Letters}\ }\textbf {\bibinfo {volume} {848}},\ \bibinfo {pages} {L15}
  (\bibinfo {year} {2017}{\natexlab{e}})}\BibitemShut {NoStop}%
\bibitem [{\citenamefont {{Coulter}}\ \emph {et~al.}(2017)\citenamefont
  {{Coulter}}, \citenamefont {{Foley}}, \citenamefont {{Kilpatrick}},
  \citenamefont {{Drout}}, \citenamefont {{Piro}}, \citenamefont {{Shappee}},
  \citenamefont {{Siebert}}, \citenamefont {{Simon}}, \citenamefont {{Ulloa}},
  \citenamefont {{Kasen}}, \citenamefont {{Madore}}, \citenamefont
  {{Murguia-Berthier}}, \citenamefont {{Pan}}, \citenamefont {{Prochaska}},
  \citenamefont {{Ramirez-Ruiz}}, \citenamefont {{Rest}},\ and\ \citenamefont
  {{Rojas-Bravo}}}]{2017Sci...358.1556C}%
  \BibitemOpen
  \bibfield  {author} {\bibinfo {author} {\bibfnamefont {D.~A.}\ \bibnamefont
  {{Coulter}}}, \bibinfo {author} {\bibfnamefont {R.~J.}\ \bibnamefont
  {{Foley}}}, \bibinfo {author} {\bibfnamefont {C.~D.}\ \bibnamefont
  {{Kilpatrick}}}, \bibinfo {author} {\bibfnamefont {M.~R.}\ \bibnamefont
  {{Drout}}}, \bibinfo {author} {\bibfnamefont {A.~L.}\ \bibnamefont {{Piro}}},
  \bibinfo {author} {\bibfnamefont {B.~J.}\ \bibnamefont {{Shappee}}}, \bibinfo
  {author} {\bibfnamefont {M.~R.}\ \bibnamefont {{Siebert}}}, \bibinfo {author}
  {\bibfnamefont {J.~D.}\ \bibnamefont {{Simon}}}, \bibinfo {author}
  {\bibfnamefont {N.}~\bibnamefont {{Ulloa}}}, \bibinfo {author} {\bibfnamefont
  {D.}~\bibnamefont {{Kasen}}}, \bibinfo {author} {\bibfnamefont {B.~F.}\
  \bibnamefont {{Madore}}}, \bibinfo {author} {\bibfnamefont {A.}~\bibnamefont
  {{Murguia-Berthier}}}, \bibinfo {author} {\bibfnamefont {Y.-C.}\ \bibnamefont
  {{Pan}}}, \bibinfo {author} {\bibfnamefont {J.~X.}\ \bibnamefont
  {{Prochaska}}}, \bibinfo {author} {\bibfnamefont {E.}~\bibnamefont
  {{Ramirez-Ruiz}}}, \bibinfo {author} {\bibfnamefont {A.}~\bibnamefont
  {{Rest}}}, \ and\ \bibinfo {author} {\bibfnamefont {C.}~\bibnamefont
  {{Rojas-Bravo}}},\ }\href {\doibase 10.1126/science.aap9811} {\bibfield
  {journal} {\bibinfo  {journal} {Science}\ }\textbf {\bibinfo {volume}
  {358}},\ \bibinfo {pages} {1556} (\bibinfo {year} {2017})},\ \Eprint
  {http://arxiv.org/abs/1710.05452} {arXiv:1710.05452 [astro-ph.HE]}
  \BibitemShut {NoStop}%
\bibitem [{\citenamefont {Lombriser}\ and\ \citenamefont
  {Taylor}(2016)}]{Lombriser:2015sxa}%
  \BibitemOpen
  \bibfield  {author} {\bibinfo {author} {\bibfnamefont {L.}~\bibnamefont
  {Lombriser}}\ and\ \bibinfo {author} {\bibfnamefont {A.}~\bibnamefont
  {Taylor}},\ }\href {\doibase 10.1088/1475-7516/2016/03/031} {\bibfield
  {journal} {\bibinfo  {journal} {JCAP}\ }\textbf {\bibinfo {volume} {1603}},\
  \bibinfo {pages} {031} (\bibinfo {year} {2016})},\ \Eprint
  {http://arxiv.org/abs/1509.08458} {arXiv:1509.08458 [astro-ph.CO]}
  \BibitemShut {NoStop}%
\bibitem [{\citenamefont {Lombriser}\ and\ \citenamefont
  {Lima}(2017)}]{Lombriser:2016yzn}%
  \BibitemOpen
  \bibfield  {author} {\bibinfo {author} {\bibfnamefont {L.}~\bibnamefont
  {Lombriser}}\ and\ \bibinfo {author} {\bibfnamefont {N.~A.}\ \bibnamefont
  {Lima}},\ }\href {\doibase 10.1016/j.physletb.2016.12.048} {\bibfield
  {journal} {\bibinfo  {journal} {Phys. Lett.}\ }\textbf {\bibinfo {volume}
  {B765}},\ \bibinfo {pages} {382} (\bibinfo {year} {2017})},\ \Eprint
  {http://arxiv.org/abs/1602.07670} {arXiv:1602.07670 [astro-ph.CO]}
  \BibitemShut {NoStop}%
\bibitem [{\citenamefont {{Baker}}\ \emph {et~al.}(2017)\citenamefont
  {{Baker}}, \citenamefont {{Bellini}}, \citenamefont {{Ferreira}},
  \citenamefont {{Lagos}}, \citenamefont {{Noller}},\ and\ \citenamefont
  {{Sawicki}}}]{2017PhRvL.119y1301B}%
  \BibitemOpen
  \bibfield  {author} {\bibinfo {author} {\bibfnamefont {T.}~\bibnamefont
  {{Baker}}}, \bibinfo {author} {\bibfnamefont {E.}~\bibnamefont {{Bellini}}},
  \bibinfo {author} {\bibfnamefont {P.~G.}\ \bibnamefont {{Ferreira}}},
  \bibinfo {author} {\bibfnamefont {M.}~\bibnamefont {{Lagos}}}, \bibinfo
  {author} {\bibfnamefont {J.}~\bibnamefont {{Noller}}}, \ and\ \bibinfo
  {author} {\bibfnamefont {I.}~\bibnamefont {{Sawicki}}},\ }\href {\doibase
  10.1103/PhysRevLett.119.251301} {\bibfield  {journal} {\bibinfo  {journal}
  {Physical Review Letters}\ }\textbf {\bibinfo {volume} {119}},\ \bibinfo
  {eid} {251301} (\bibinfo {year} {2017})},\ \Eprint
  {http://arxiv.org/abs/1710.06394} {arXiv:1710.06394} \BibitemShut {NoStop}%
\bibitem [{\citenamefont {Creminelli}\ and\ \citenamefont
  {Vernizzi}(2017)}]{Creminelli:2017sry}%
  \BibitemOpen
  \bibfield  {author} {\bibinfo {author} {\bibfnamefont {P.}~\bibnamefont
  {Creminelli}}\ and\ \bibinfo {author} {\bibfnamefont {F.}~\bibnamefont
  {Vernizzi}},\ }\href {\doibase 10.1103/PhysRevLett.119.251302} {\bibfield
  {journal} {\bibinfo  {journal} {Phys. Rev. Lett.}\ }\textbf {\bibinfo
  {volume} {119}},\ \bibinfo {pages} {251302} (\bibinfo {year} {2017})},\
  \Eprint {http://arxiv.org/abs/1710.05877} {arXiv:1710.05877 [astro-ph.CO]}
  \BibitemShut {NoStop}%
\bibitem [{\citenamefont {Sakstein}\ and\ \citenamefont
  {Jain}(2017)}]{Sakstein:2017xjx}%
  \BibitemOpen
  \bibfield  {author} {\bibinfo {author} {\bibfnamefont {J.}~\bibnamefont
  {Sakstein}}\ and\ \bibinfo {author} {\bibfnamefont {B.}~\bibnamefont
  {Jain}},\ }\href {\doibase 10.1103/PhysRevLett.119.251303} {\bibfield
  {journal} {\bibinfo  {journal} {Phys. Rev. Lett.}\ }\textbf {\bibinfo
  {volume} {119}},\ \bibinfo {pages} {251303} (\bibinfo {year} {2017})},\
  \Eprint {http://arxiv.org/abs/1710.05893} {arXiv:1710.05893 [astro-ph.CO]}
  \BibitemShut {NoStop}%
\bibitem [{\citenamefont {Ezquiaga}\ and\ \citenamefont
  {Zumalacarregui}(2017)}]{Ezquiaga:2017ekz}%
  \BibitemOpen
  \bibfield  {author} {\bibinfo {author} {\bibfnamefont {J.~M.}\ \bibnamefont
  {Ezquiaga}}\ and\ \bibinfo {author} {\bibfnamefont {M.}~\bibnamefont
  {Zumalacarregui}},\ }\href {\doibase 10.1103/PhysRevLett.119.251304}
  {\bibfield  {journal} {\bibinfo  {journal} {Phys. Rev. Lett.}\ }\textbf
  {\bibinfo {volume} {119}},\ \bibinfo {pages} {251304} (\bibinfo {year}
  {2017})},\ \Eprint {http://arxiv.org/abs/1710.05901} {arXiv:1710.05901
  [astro-ph.CO]} \BibitemShut {NoStop}%
\bibitem [{\citenamefont {Bettoni}\ and\ \citenamefont
  {Liberati}(2013)}]{Bettoni:2013diz}%
  \BibitemOpen
  \bibfield  {author} {\bibinfo {author} {\bibfnamefont {D.}~\bibnamefont
  {Bettoni}}\ and\ \bibinfo {author} {\bibfnamefont {S.}~\bibnamefont
  {Liberati}},\ }\href {\doibase 10.1103/PhysRevD.88.084020} {\bibfield
  {journal} {\bibinfo  {journal} {Phys. Rev.}\ }\textbf {\bibinfo {volume}
  {D88}},\ \bibinfo {pages} {084020} (\bibinfo {year} {2013})},\ \Eprint
  {http://arxiv.org/abs/1306.6724} {arXiv:1306.6724 [gr-qc]} \BibitemShut
  {NoStop}%
\bibitem [{\citenamefont {Yunes}\ and\ \citenamefont
  {Siemens}(2013)}]{Yunes:2013dva}%
  \BibitemOpen
  \bibfield  {author} {\bibinfo {author} {\bibfnamefont {N.}~\bibnamefont
  {Yunes}}\ and\ \bibinfo {author} {\bibfnamefont {X.}~\bibnamefont
  {Siemens}},\ }\href {\doibase 10.12942/lrr-2013-9} {\bibfield  {journal}
  {\bibinfo  {journal} {Living Rev. Rel.}\ }\textbf {\bibinfo {volume} {16}},\
  \bibinfo {pages} {9} (\bibinfo {year} {2013})},\ \Eprint
  {http://arxiv.org/abs/1304.3473} {arXiv:1304.3473 [gr-qc]} \BibitemShut
  {NoStop}%
\bibitem [{\citenamefont {Gong}\ and\ \citenamefont
  {Hou}(2018)}]{Gong:2017bru}%
  \BibitemOpen
  \bibfield  {author} {\bibinfo {author} {\bibfnamefont {Y.}~\bibnamefont
  {Gong}}\ and\ \bibinfo {author} {\bibfnamefont {S.}~\bibnamefont {Hou}},\
  }\bibfield  {booktitle} {\emph {\bibinfo {booktitle} {{Proceedings, 13th
  International Conference on Gravitation, Astrophysics and Cosmology and 15th
  Italian-Korean Symposium on Relativistic Astrophysics (IK15): Seoul, Korea,
  July 3-7, 2017}}},\ }\href {\doibase 10.1051/epjconf/201816801003} {\bibfield
   {journal} {\bibinfo  {journal} {EPJ Web Conf.}\ }\textbf {\bibinfo {volume}
  {168}},\ \bibinfo {pages} {01003} (\bibinfo {year} {2018})},\ \Eprint
  {http://arxiv.org/abs/1709.03313} {arXiv:1709.03313 [gr-qc]} \BibitemShut
  {NoStop}%
\bibitem [{\citenamefont {{Hawking}}(1972)}]{1972CMaPh..25..152H}%
  \BibitemOpen
  \bibfield  {author} {\bibinfo {author} {\bibfnamefont {S.~W.}\ \bibnamefont
  {{Hawking}}},\ }\href {\doibase 10.1007/BF01877517} {\bibfield  {journal}
  {\bibinfo  {journal} {Communications in Mathematical Physics}\ }\textbf
  {\bibinfo {volume} {25}},\ \bibinfo {pages} {152} (\bibinfo {year}
  {1972})}\BibitemShut {NoStop}%
\bibitem [{\citenamefont {{Leaver}}(1985)}]{1985RSPSA.402..285L}%
  \BibitemOpen
  \bibfield  {author} {\bibinfo {author} {\bibfnamefont {E.~W.}\ \bibnamefont
  {{Leaver}}},\ }\href {\doibase 10.1098/rspa.1985.0119} {\bibfield  {journal}
  {\bibinfo  {journal} {Proceedings of the Royal Society of London Series A}\
  }\textbf {\bibinfo {volume} {402}},\ \bibinfo {pages} {285} (\bibinfo {year}
  {1985})}\BibitemShut {NoStop}%
\bibitem [{\citenamefont {{Teukolsky}}(1973)}]{1973ApJ...185..635T}%
  \BibitemOpen
  \bibfield  {author} {\bibinfo {author} {\bibfnamefont {S.~A.}\ \bibnamefont
  {{Teukolsky}}},\ }\href {\doibase 10.1086/152444} {\bibfield  {journal}
  {\bibinfo  {journal} {\apj}\ }\textbf {\bibinfo {volume} {185}},\ \bibinfo
  {pages} {635} (\bibinfo {year} {1973})}\BibitemShut {NoStop}%
\bibitem [{\citenamefont {{Pani}}\ \emph {et~al.}(2012)\citenamefont {{Pani}},
  \citenamefont {{Cardoso}}, \citenamefont {{Gualtieri}}, \citenamefont
  {{Berti}},\ and\ \citenamefont {{Ishibashi}}}]{2012PhRvD..86j4017P}%
  \BibitemOpen
  \bibfield  {author} {\bibinfo {author} {\bibfnamefont {P.}~\bibnamefont
  {{Pani}}}, \bibinfo {author} {\bibfnamefont {V.}~\bibnamefont {{Cardoso}}},
  \bibinfo {author} {\bibfnamefont {L.}~\bibnamefont {{Gualtieri}}}, \bibinfo
  {author} {\bibfnamefont {E.}~\bibnamefont {{Berti}}}, \ and\ \bibinfo
  {author} {\bibfnamefont {A.}~\bibnamefont {{Ishibashi}}},\ }\href {\doibase
  10.1103/PhysRevD.86.104017} {\bibfield  {journal} {\bibinfo  {journal}
  {\prd}\ }\textbf {\bibinfo {volume} {86}},\ \bibinfo {eid} {104017} (\bibinfo
  {year} {2012})},\ \Eprint {http://arxiv.org/abs/1209.0773} {arXiv:1209.0773
  [gr-qc]} \BibitemShut {NoStop}%
\bibitem [{\citenamefont {{Teo}}(2003)}]{2003GReGr..35.1909T}%
  \BibitemOpen
  \bibfield  {author} {\bibinfo {author} {\bibfnamefont {E.}~\bibnamefont
  {{Teo}}},\ }\href {\doibase 10.1023/A:1026286607562} {\bibfield  {journal}
  {\bibinfo  {journal} {General Relativity and Gravitation}\ }\textbf {\bibinfo
  {volume} {35}},\ \bibinfo {pages} {1909} (\bibinfo {year}
  {2003})}\BibitemShut {NoStop}%
\end{thebibliography}%

\end{document}